\documentclass[sigconf, nonacm]{acmart}
\usepackage{enumitem}
\usepackage{xcolor} % Required for color support
\usepackage{pifont} % For checkmark and cross symbols
\usepackage{colortbl} % Required for \rowcolor in tables
\usepackage{makecell}
\usepackage{mathtools} 
\usepackage{tikz}
\usepackage{pgfplots}
\usepackage{pgfplotstable}
\usepackage{subcaption}
\usepackage{hyperref}
\usepackage{flushend}
\pgfplotsset{compat=1.16}

\usepackage{tcolorbox}     % 画带标题的彩色盒子
\usepackage{enumitem}      % 控制 itemize, enumerate 的边距和间距
\usepackage{multirow}

\usepackage[linesnumbered,ruled,vlined]{algorithm2e} %[ruled,vlined]{
\usepackage{algpseudocode}
\usepackage{amsmath}
\SetKwInput{KwInput}{Input}
\SetKwInput{KwOutput}{Output}
\SetKwFor{ParFor}{\textbf{parallel for}}{}{end}
\newcommand{\bluecomment}[1]{\textcolor{blue}{\texttt{/* #1 */}}}

\usepgfplotslibrary{groupplots}
\pgfplotsset{compat=1.18}   
\usetikzlibrary{patterns,shapes.geometric,calc,positioning,arrows.meta}

\newtheoremstyle{examplestyle} % name
    {0pt} % Space above
    {0pt} % Space below
    {\itshape} % Body font
    {} % Indent amount
    {\bfseries} % Theorem head font
    {.} % Punctuation after theorem head
    {.1em} % Space after theorem head
    {} % Theorem head spec (can be left empty, meaning 'normal')
\theoremstyle{examplestyle}
\newtheorem{example}{Example}
\newtheoremstyle{definitionstyle} % name of the style
    {0pt} % Space above
    {0pt} % Space below
    {\itshape} % Body font
    {} % Indent amount
    {\bfseries} % Theorem head font (bold and small caps)
    {.} % Punctuation after theorem head
    {.3em} % Space after theorem head
    {} % Theorem head spec (can be left empty, meaning 'normal')
\theoremstyle{definitionstyle}
\newtheorem{definition}{Definition}

\usepackage{tcolorbox}
\newcommand\vldbdoi{XX.XX/XXX.XX}
\newcommand\vldbpages{XXX-XXX}
\newcommand\vldbvolume{19}
\newcommand\vldbissue{1}
\newcommand\vldbyear{2026}
\newcommand\vldbauthors{\authors}
\newcommand\vldbtitle{\shorttitle}
\newcommand\vldbavailabilityurl{https://github.com/hyLiu1994/VISTA}
\newcommand\vldbpagestyle{plain}

\begin{document}
\title{VISTA: Knowledge-Driven Vessel Trajectory Imputation with Repair Provenance}
%%
%% The "author" command and its associated commands are used to define the authors and their affiliations.

\author{Hengyu Liu$^1$ \quad Tianyi Li$^{1\dagger}$ \quad Haoyu Wang$^2$ \quad Kristian Torp$^1$ \quad Tiancheng Zhang$^2$ \quad Yushuai Li$^1$  \quad Christian S. Jensen$^1$}
\affiliation{%
  \institution{$^1$Department of Computer Science, Aalborg University, Denmark\\
  $^2$School of Computer Science and Engineering, Northeastern University, Shenyang, China}
  \country{} 
  $^1$\{heli, tianyi, torp, yusli, csj\}@cs.aau.dk, $^2$haoyu4260@gmail.com, tczhang@mail.neu.edu.cn}

%%
%% The abstract is a short summary of the work to be presented in the article.
\begin{abstract}
Repairing incomplete trajectory data is essential for downstream spatio-temporal applications. Yet, existing repair methods focus solely on reconstruction without documenting the reasoning behind repair decisions, undermining trust in safety-critical applications where repaired trajectories affect operational decisions, such as in maritime anomaly detection and route planning. We introduce \emph{repair provenance}---structured, queryable metadata that documents the full reasoning chain behind each repair---which transforms imputation from pure data recovery into a task that supports downstream decision-making.
We propose \textsf{VISTA} (knowledge-dri\underline{\textbf{v}}en \underline{\textbf{i}}nterpretable ve\underline{\textbf{s}}sel \underline{\textbf{t}}rajectory imput\underline{\textbf{a}}tion), a framework that reliably equips repaired trajectories with repair provenance by grounding LLM reasoning in data-verified knowledge.
Specifically, we formalize Structured Data-derived Knowledge (SDK), a knowledge model whose data-verifiable components can be validated against real data and used to anchor and constrain LLM-generated explanations.
We organize SDK in a Structured Data-derived Knowledge Graph (SD-KG) and establish a data--knowledge--data loop for extraction, validation, and incremental maintenance over large-scale AIS data.
A workflow management layer with parallel scheduling, fault tolerance, and redundancy control ensures consistent and efficient end-to-end processing.
Experiments on two large-scale AIS datasets show that \textsf{VISTA} achieves state-of-the-art accuracy, improving over baselines by 5\%--91\% and reducing inference time by 51\%--93\%, while producing repair provenance, whose interpretability is further validated through a case study and an interactive demo system.
\end{abstract}
\maketitle

\pagestyle{\vldbpagestyle}
\begingroup\small\noindent\raggedright\textbf{PVLDB Reference Format:}\\
\vldbauthors. \vldbtitle. PVLDB, \vldbvolume(\vldbissue): \vldbpages, \vldbyear.\\
\href{https://doi.org/\vldbdoi}{doi:\vldbdoi}
\endgroup
\begingroup
\renewcommand\thefootnote{}\footnote{\noindent
$^{\dagger}$ Tianyi Li is the corresponding author.\\
This work is licensed under the Creative Commons BY-NC-ND 4.0 International License. Visit \url{https://creativecommons.org/licenses/by-nc-nd/4.0/} to view a copy of this license. For any use beyond those covered by this license, obtain permission by emailing \href{mailto:info@vldb.org}{info@vldb.org}. Copyright is held by the owner/author(s). Publication rights licensed to the VLDB Endowment. \\
\raggedright Proceedings of the VLDB Endowment, Vol. \vldbvolume, No. \vldbissue\ %
ISSN 2150-8097. \\
\href{https://doi.org/\vldbdoi}{doi:\vldbdoi} \\
}\addtocounter{footnote}{-1}\endgroup

\ifdefempty{\vldbavailabilityurl}{}{
\vspace{.3cm}
\begingroup\small\noindent\raggedright\textbf{PVLDB Artifact Availability:}\\
The source code, data, and/or other artifacts have been made available at \url{https://github.com/hyLiu1994/VISTA}.
\endgroup
}

\section{Introduction}
Repairing incomplete trajectory data is essential for downstream applications such as route planning, traffic monitoring, and anomaly detection~\cite{fan2008conditional,rekatsinas2017holoclean,bank_methodological_2023,song_screen_2015,rezig2021horizon}.
While existing repair methods---whether rule-based~\cite{widyantara_improvement_2023,wang_kinematic_2023}, learning-based~\cite{jiang2024stmgf,syed2023cnn,liu2025mhgin}, or LLM-driven~\cite{narayan2022can,musleh2023kamel}---differ in transparency, none produces structured, queryable metadata that documents the reasoning behind repair decisions.
In safety-critical domains such as maritime safety via the Automatic Identification System (AIS)~\cite{tu2018exploiting}, where repaired data affects operational decisions, the absence of such metadata undermines the trustworthiness of the entire analytical pipeline~\cite{cheney2009provenance}.

% P2: introduce repair provenance
Classical data provenance~\cite{buneman2001and,cheney2009provenance} tracks how query results are derived from source tuples, but does not capture the reasoning behind repair decisions. To bridge this gap, we introduce \textbf{repair provenance}, structured and queryable metadata that records the full reasoning chain behind each repaired value through three components:
i)~\emph{what} behavior pattern is estimated for the missing segment, supported by statistical evidence;
ii)~\emph{how} the segment is imputed, justified by the inferred pattern; and
iii)~\emph{why} the behavior occurred, grounded in a human-readable causal explanation.

\begin{figure}[!t]
    \centering
    \includegraphics[width=1\linewidth]{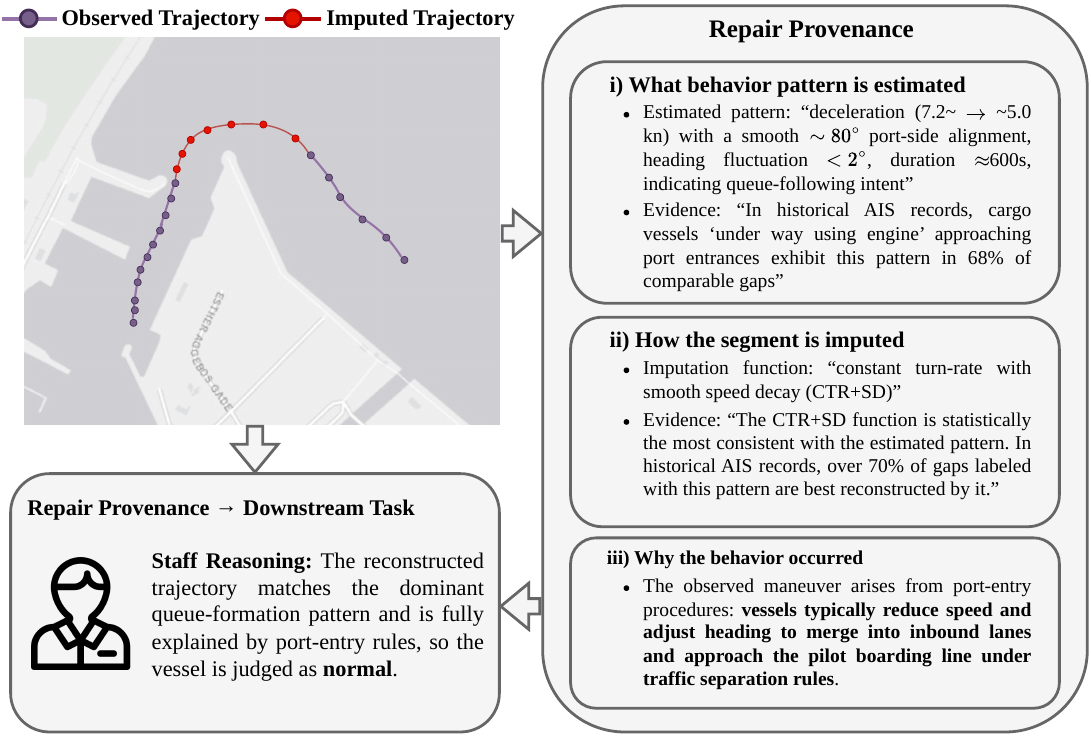}
    \vspace{-6mm}
    \caption{An imputed trajectory with repair provenance. Upper-left: observed (purple) and imputed (red) segments near a port. Right: provenance components---behavior pattern, imputation function, and causal explanation. Lower-left: provenance-supported anomaly detection.}
    \label{fig:intro-fi-1}
    \vspace{-7mm}
\end{figure}

\begin{example}\label{exp:1}
A cargo ship's AIS messages (status: ``under way using engine'') are lost for about ten minutes near a busy port. The missing sub-trajectory (red curve in Figure~\ref{fig:intro-fi-1}) is reconstructed as a smooth curve between the observed segments (purple points). The accompanying repair provenance (right-side panels) encompasses three components:

\noindent \textbf{i) What behavior pattern is estimated.} The estimated pattern is a deceleration from 7.2 to 5.0kn with a smooth ${\sim}80^{\circ}$ port-side alignment, heading fluctuation ${<}2^{\circ}$, and duration ${\approx}600$s, indicating queue-following intent. In historical AIS data, cargo vessels approaching port entrances exhibit this pattern in 68\% of comparable gaps.

\noindent \textbf{ii) How the segment is imputed.} The selected imputation function, constant turn-rate with smooth speed decay (CTR+SD), is statistically the most consistent with the estimated pattern, accounting for over 70\% of successful reconstructions under this pattern.

\noindent \textbf{iii) Why the behavior occurred.} The behavior reflects port-entry procedures: vessels reduce speed and adjust heading to merge into inbound lanes and approach the pilot boarding line under traffic separation rules. This causal explanation distinguishes routine port-entry maneuvers from anomalous behaviors such as evasive steering or unauthorized anchoring.

As illustrated in the lower-left panel, this provenance directly supports downstream anomaly detection, enabling the operator to judge the vessel as non-anomalous.
\end{example}

This illustrates how \textbf{repair provenance can transform trajectory imputation from a pure data reconstruction task to one that better supports downstream decision-making.}
Large language models (LLMs)~\cite{liang_foundation_2025,zhang_large_2024,liang_kag_2025}, with their capacity for contextual reasoning, knowledge integration, and text generation, offer the first technical means to produce such provenance automatically, yet leveraging them is challenging.

\noindent \textbf{\textit{C1: How to define a knowledge model that grounds and constrains LLM reasoning?}}
LLM outputs are unreliable---they may hallucinate behavior patterns, generate invalid imputation functions, or fabricate statistical evidence~\cite{shankar2024spade,agrawal2024kghallucination}. To produce trustworthy repair provenance, a structured knowledge model is needed whose data-verifiable components---behavior patterns, imputation functions, and their statistical associations---can be validated against real data and used to anchor and constrain LLM-generated explanations. No existing formalism provides this: classical provenance models~\cite{buneman2001and,cheney2009provenance,green2007provenance} track query derivations but not repair reasoning, while data repair systems~\cite{rekatsinas2017holoclean,geerts2013llunatic,rammelaere2018xplode} either lack explanations or produce unverified, unstructured annotations.

\noindent \textbf{\textit{C2: How to construct and maintain a data-verified knowledge base?}}
The knowledge model above provides a structural blueprint, but its components must be populated with empirically grounded content---behavior patterns statistically supported by observed data, imputation functions that execute correctly on real trajectories, and associations backed by verified evidence~\cite{yang_comprehensive_2025,fernandez2023large}. This requires extracting structured knowledge from massive, noisy trajectory records, validating each component through execution and statistical testing, and updating the knowledge base as new data arrives~\cite{hogan2021knowledge,liang_kag_2025}. Only then can it reliably ground and constrain LLM reasoning for provenance generation.

\noindent \textbf{\textit{C3: How to impute trajectories with repair provenance reliably at scale?}}
The full pipeline---from knowledge extraction through trajectory imputation to provenance generation---involves numerous LLM inference calls that may vary across runs, may time out, or may produce redundant artifacts~\cite{ahn2025prompt,tan2025too,agrawal2024taming}. Moreover, the steps are interdependent: a failed pattern extraction affects subsequent imputation and provenance steps. An execution framework must therefore ensure coordinated parallel scheduling, fault tolerance, and redundancy control to achieve consistent end-to-end \mbox{processing}.

\looseness=-1
To address these challenges, we propose \textsf{VISTA}, a knowledge-dri\underline{\textbf{v}}en \underline{\textbf{i}}nterpretable ve\underline{\textbf{s}}sel \underline{\textbf{t}}rajectory imput\underline{\textbf{a}}tion framework that equips repaired trajectories with structured repair provenance.

\noindent \textbf{\textit{To address C1}}, we propose Structured Data-derived Knowledge (SDK), a knowledge model that organizes data-verifiable knowledge, vessel attributes, behavior patterns, imputation functions, and their statistical associations, in a relational schema suitable for storage, querying, and validation. Validated SDK serves as grounded context for steering LLM reasoning towards correct causal explanations rather than hallucinated ones. Together, SDK-derived facts and LLM-generated explanations form complete repair provenance records.

\noindent \textbf{\textit{To address C2}}, we establish a data--knowledge--data loop with a Structured Data-derived Knowledge Graph (SD-KG) at its core. Going from data to knowledge, \textsf{VISTA} extracts knowledge units from raw trajectory records and validates each unit through execution and statistical testing before admitting it into the SD-KG. Going from knowledge to data, validated SD-KG guides trajectory imputation and provenance generation. The repaired data is then fed back into the SD-KG, closing the loop and incrementally enriching the knowledge base with every repaired trajectory.

\noindent \textbf{\textit{To address C3}}, we design a workflow management layer that coordinates the full end-to-end pipeline. It employs parallel task scheduling, fault tolerance, and redundancy control to handle the instability inherent in large-scale LLM inference. This ensures that failures at individual steps do not cascade and that the end-to-end process remains consistent and efficient.

To validate the generated repair provenance, we provide a case study that qualitatively examines its interpretability and develop \texttt{CLEAR}~\cite{liu2025clear}, an interactive demo system that enables non-expert users to explore, complete, and annotate vessel trajectories through a knowledge-graph-driven interface. The main contributions are summarized as follows:

\begin{itemize}[itemsep=1pt, leftmargin=10pt]
\item We propose \textsf{VISTA}, a knowledge-driven interpretable vessel trajectory imputation framework that reliably equips repaired trajectories with structured repair provenance at scale by grounding LLM reasoning in data-verified knowledge.

\item We formalize Structured Data-derived Knowledge and organize it in a Structured Data-derived Knowledge Graph, establishing a data--knowledge--data loop for extraction, validation, incremental maintenance, and context-aware querying.

\item We design a workflow management layer featuring parallel scheduling, fault tolerance, and redundancy control to enable consistent and efficient processing of large-scale AIS data.

\item Experiments on two AIS datasets show that \textsf{VISTA} is capable of state-of-the-art accuracy, outperforming existing baselines by 5\%--91\% and reducing inference time by 51\%--93\%, while producing repair provenance.
\end{itemize}

The rest of the paper is structured as follows:
Section~\ref{sec:related_work} reviews related work.
Section~\ref{sec:preliminaries} covers notation, the knowledge model, and the problem formulation.
Section~\ref{sec:methodology} presents \textsf{VISTA}.
Section~\ref{sec:experiment} reports experimental results, and
Section~\ref{sec:conclusion} concludes the paper.

\section{Related Work} \label{sec:related_work}

Leveraging LLMs to produce repair provenance for trajectory imputation touches three areas: trajectory imputation, data provenance and repair explanation, and knowledge-enhanced LLM reasoning.

\noindent \textbf{Trajectory Imputation.}
Trajectory imputation has been studied using four approaches, each advancing reconstruction accuracy while leaving repair decisions undocumented.
\emph{Rule-based} methods repair data by enforcing declarative constraints---functional dependencies (FDs), such as conditional FDs (CFDs)~\cite{fan2008conditional}, and denial constraints~\cite{chu2013discovering}---or domain-specific speed and acceleration bounds~\cite{song_screen_2015}, with scalable systems such as Horizon~\cite{rezig2021horizon} enabling FD-based cleaning over large datasets. These methods require manual rule specification and do not record why a particular repair value was chosen over alternatives.
\emph{Interpolation-based} methods fit predefined mathematical models to observed data, such as linear and spline interpolation~\cite{widyantara_improvement_2023,zaman_interpolation-based_2023}. These methods are used widely as preprocessing steps in trajectory analysis pipelines due to their simplicity and low computational cost. However, they assume smooth dynamics between the observed points and offer limited adaptability to complex movement patterns such as abrupt turns or speed changes.
\emph{Deep learning-based} methods learn spatio-temporal patterns directly from large-scale data. General-purpose methods such as GAIN~\cite{yoon2018gain}, SAITS~\cite{du_saits_2023}, CSDI~\cite{tashiro2021csdi}, and DiffPuter~\cite{zhang2025diffputer} handle missing multivariate data through adversarial training, self-attention, or diffusion models, while BERT-based methods such as KAMEL~\cite{musleh2023kamel} and TrajBERT~\cite{si2023trajbert} adapt masked-token pre-training to spatial sequences. These methods achieve state-of-the-art reconstruction accuracy but remain black boxes with no transparency into individual repair decisions.
\emph{LLM-based} methods leverage the contextual reasoning and text generation capabilities of LLMs for data management tasks. For data cleaning, foundation models have been evaluated on imputation and error detection~\cite{narayan2022can}, and RetClean~\cite{retclean2024} combines retrieval from data lakes with LLM inference to repair dirty data. In the maritime domain, LLMs have been applied to related trajectory tasks such as prediction and anomaly detection~\cite{yu2025multi,park2025ais}, but not yet to imputation with provenance. While LLM-based methods can produce free-text justifications alongside their outputs, such explanations are post-hoc rationalizations rather than faithful accounts of the repair process~\cite{xu2025trajectory,li2024llm}, and they are neither structured nor queryable.
Across all four approaches, no method produces structured repair provenance alongside the imputed values.

\noindent \textbf{Data Provenance and Repair Explanation.}
Classical data provenance tracks how query results derive from source tuples. Thus, why- and where-prove\-nance~\cite{cheney2009provenance,buneman2001and} identify contributing source tuples and pinpoint copied attribute values, respectively, while how-provenance~\cite{green2007provenance} captures computation structure via semiring annotations. Data lineage~\cite{cui2000tracing} and provenance-aware systems such as Perm~\cite{glavic2009perm} further support debugging and auditing through query rewriting. Such provenance is structured and queryable via standard SQL. However, classical provenance tracks data \emph{derivation}---how query results are computed from existing data---rather than data \emph{repair reasoning}---why a missing value was replaced with a specific alternative over others.

Several systems aim to justify individual repair decisions. NAD\-EEF~\cite{dallachiesa2013nadeef} stores violation and repair metadata in relational tables, recording which rule triggered each change, but does not capture why a particular replacement value was chosen over alternatives. LLUNATIC~\cite{geerts2013llunatic} derives repairs through chase-based reasoning with formal guarantees. Yet, its justification is implicit in the chase trace rather than materialized as queryable metadata. HoloClean~\cite{rekatsinas2017holoclean} outputs per-value confidence scores through probabilistic inference, but these scores reflect global model weights rather than per-value signal attributions. XPlode~\cite{rammelaere2018xplode} identifies the minimal CFD set, justifying a group of repairs, producing constraint-level, rather than per-value, explanations. Complementary efforts address related goals: CAPE~\cite{miao2019cape} explains query outliers via pattern-based counterbalances; counterfactual methods~\cite{galhotra2021explaining} and Gopher~\cite{zhu2022gopher} reveal ML decision boundaries and bias sources; self-supervised methods~\cite{peng2022interpretable} learn interpretable repair rules; and in-database imputation~\cite{perini2024indb} integrates MICE within a DBMS.

In summary, classical provenance is structured and queryable but does not address repair reasoning. Existing repair explanation methods provide partial justificati\-on---audit logs, confidence scores, or constraint references---but none produces a unified, queryable record of the full reasoning chain behind each repair.

\noindent \textbf{Knowledge-Enhanced LLM Reasoning.}
Knowledge graphs provide structured, queryable representations of domain facts~\cite{hogan2021knowledge} that have recently been leveraged to ground LLM reasoning and reduce hallucination~\cite{pan2024unifying,agrawal2024kghallucination,yang_comprehensive_2025}. Pan et al.~\cite{pan2024unifying} provide a roadmap categorizing approaches into KG-enhanced LLMs, LLM-augmented KGs, and synergized LLM+KG systems. KAG~\cite{liang_kag_2025} combines logical-form-guided retrieval with vector search to enhance LLM reasoning in professional domains. On the retrieval side, retrieval-augmented generation (RAG)~\cite{lewis2020rag} anchors LLM outputs in external evidence, and graph-structured extensions~\cite{peng2024graphrag} further incorporate relational knowledge into the retrieval process.

However, existing methods are designed primarily for open-domain question answering or text generation; the knowledge they retrieve is typically accepted without validation against the data. At scale, this absence of validation is compounded by LLM-specific reliability challenges, prompt inconsistency~\cite{ahn2025prompt}, self-consistent errors~\cite{tan2025too}, and resource contention~\cite{agrawal2024taming}, that can degrade output quality. No existing studies validate knowledge against the data before using it to guide LLM reasoning; nor do they provide pipeline-level mechanisms to ensure reliable execution at scale.

\section{Preliminaries} \label{sec:preliminaries}
\subsection{Data and Notation}
\begin{definition}[\textbf{AIS Record}] \label{def-1}
An AIS record $x = (\iota, \lambda, \phi, \tau, \psi, \theta, s, \eta, \chi,\\ d, \ell, \beta, \kappa )$ has multiple attributes.
At its \textbf{core}, an AIS record is anchored by a vessel identifier and spatio-temporal attributes $(\iota, \lambda, \phi, \tau)$, representing vessel identifier, longitude, latitude, and timestamp, identifying a vessel and giving its geographic position at a particular time. Other attributes act as \textbf{auxiliary \mbox{descriptors}}:  
\begin{itemize}[itemsep=2pt, leftmargin=10pt]
    \item \textbf{Kinematic:} Heading angle $\psi$, course over ground $\theta$, and speed over ground $s$ describe the vessel’s movement.  
    \item \textbf{Status-related:} Navigation status $\eta$, hazardous cargo type $\chi$, and draught $d$ provide operational and safety context.  
    \item \textbf{Static:} Vessel length $\ell$, width $\beta$, and type $\kappa$ characterize vessel geometry and category. 
\end{itemize} 
\end{definition}

\begin{definition}[\textbf{Vessel-specific AIS Record Sequence}] \label{def-2}
Given a vessel $\iota$, an AIS record sequence $\mathbf{X}_{\iota} = \langle x_1, x_2, \dots, x_T \rangle$ is a time-ordered sequence, where each $x_t$ is as defined in Definition~\ref{def-1}. The timestamps $\tau$ satisfy $\tau_{t} < \tau_{t+1}$ for all $t \in [1, T-1]$. 
\end{definition}

A vessel-specific AIS record sequence $\mathbf{X}_{\iota}$ captures the spatio-temporal trajectory of vessel $\iota$, together with its auxiliary descriptors (kinematic, status-related, and static attributes) across the time period $[1, T]$. Given a collection of vessels, the entire AIS data is represented as:
\begin{equation}
\mathcal{X} = \{\mathbf{X}_{\iota_1}, \mathbf{X}_{\iota_2}, \dots, \mathbf{X}_{\iota_N}\},  
\end{equation}
where each $\mathbf{X}_{\iota_i}$ denotes the AIS record sequence of vessel $\iota_i$, and $N$ is the total number of vessels. 

\subsection{Structured Data-derived Knowledge}
\begin{figure}[!tbp]
\centering
\includegraphics[width=0.9\linewidth]{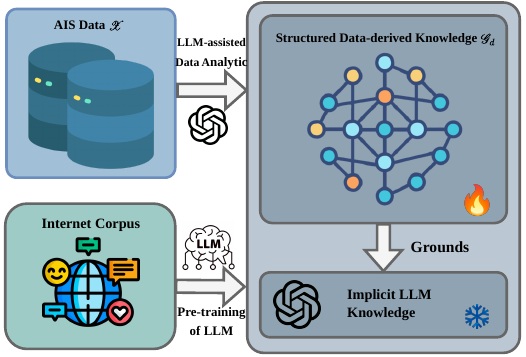}
\caption{Role of SDK in grounding LLM reasoning for trajectory imputation with repair provenance.}
\label{fig:node_relation}
\vspace{-3mm}
\end{figure}

LLM outputs are prone to hallucination and fabrication~\cite{shankar2024spade,agrawal2024kghallucination}. To produce trustworthy repair provenance, \textsf{VISTA} grounds LLM reasoning in \emph{Structured Data-derived Knowledge (SDK)}: a set of data-verifiable components---vessel attributes, behavior patterns, imputation functions, and their statistical associa\-tions---that are extracted from AIS data $\mathcal{X}$, validated through execution and statistical testing. As shown in Figure~\ref{fig:node_relation}, this data-derived and continuously updated knowledge complements an LLM's implicit knowledge acquired through pre-training on internet corpora and fixed at deployment. During imputation, validated SDK serves as grounded context that steers LLM reasoning toward faithful causal explanations, while the LLM contributes contextual reasoning and text generation capabilities.

To enable efficient storage, querying, and retrieval of SDK, we introduce the \emph{Structured Data-derived Knowledge Graph (SD-KG)}. 

\begin{definition}[\textbf{Structured Data-derived Knowledge Graph}] 
An SD-KG is a graph $\mathcal{G}_d = (\mathcal{V}_d, \mathcal{E}_d)$, distilled from AIS data $\mathcal{X}$. The node set $\mathcal{V}_d = \mathcal{V}_s \cup \mathcal{V}_b \cup \mathcal{V}_f$ is composed as follows:  

\begin{itemize}[itemsep=2pt, leftmargin=12pt]
    \item \textbf{Static Attribute Nodes} $\mathcal{V}_s = \mathcal{V}_{\iota} \cup \mathcal{V}_{\eta} \cup \mathcal{V}_{\chi} \cup \mathcal{V}_{d} \cup \mathcal{V}_{\ell} \cup \mathcal{V}_{\beta} \cup \mathcal{V}_{\kappa} \cup \mathcal{V}_{\sigma}$, mostly derived from the static and status-related attributes of AIS records (Definition~\ref{def-1}). For each attribute, a corresponding node subset is constructed, where each node represents one possible value of that attribute.  
    $\mathcal{V}_{\sigma}$ differs from other sets in $\mathcal{V}_s$ as it is derived from core attributes and represents spatial context categories such as shipping lanes, ports, or anchorages.

    \item \textbf{Behavior Pattern Nodes} $\mathcal{V}_b$ representing characteristic behavior patterns extracted from AIS data.  
    Each node $v_b \in \mathcal{V}_b$ is defined as a tuple $v_b = (p^s, p^{\theta}, p^{\psi}, p^i, p^{\tau})$, where $p^s$ is the speed pattern, $p^{\theta}$ the course pattern, $p^{\psi}$ the heading pattern, $p^i$ the navigation intent, and $p^{\tau}$ the duration. These patterns are primarily inferred from the kinematic and core attributes.
    \item \textbf{Imputation Function Nodes} $\mathcal{V}_f$ representing available imputation methods. Each node $v_f \in \mathcal{V}_f$ is defined as a tuple $v_f = ( f, d(f) )$, where $f$ is an imputation function and $d(f)$ is its descriptive information. These nodes are primarily linked to the types of behavior patterns they can address, which are inferred from the core attributes of AIS records.
\end{itemize}  
The edge set $\mathcal{E}_d = \mathcal{E}_{sb} \cup \mathcal{E}_{bf}$ consists of two types of edges:  
\begin{itemize}[itemsep=2pt, leftmargin=12pt]
    \item \textbf{Static–Behavior edges} $\mathcal{E}_{sb} \subseteq \mathcal{V}_s \times \mathcal{V}_b$:  
    connecting static attribute nodes and behavior pattern nodes.  
    For an edge $e = (v_s, v_b, w) \\ \in \mathcal{E}_{sb}$, the weight $w$ indicates how frequently the static attribute value $v_s$ co-occurs with the behavior pattern $v_b$ in $\mathcal{X}$.
    \item \textbf{Behavior–Function edges} $\mathcal{E}_{bf} \subseteq \mathcal{V}_b \times \mathcal{V}_f$:  
    connecting behavior pattern nodes and imputation function nodes.  
    For an edge $e = (v_b, v_f, w) \in \mathcal{E}_{bf}$, the weight $w$ is the frequency with which the imputation method $v_f$ successfully imputes trajectories exhibiting the behavior pattern $v_b$.
\end{itemize}  
\end{definition}

\begin{figure}[!tbp]
\centering
\includegraphics[width=1.03\linewidth]{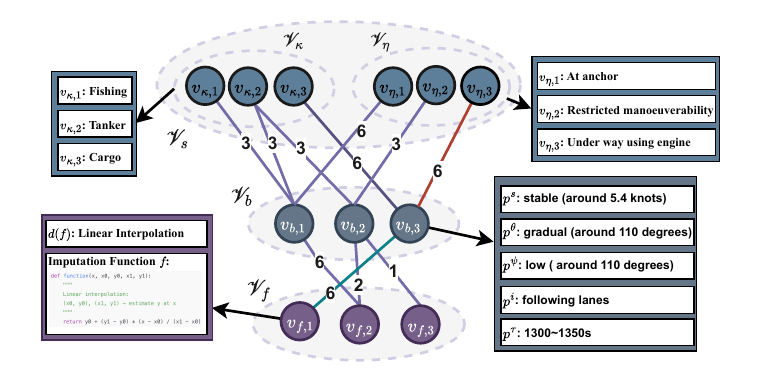}
\caption{Example of SD-KG $\mathcal{G}_d$.}
\label{fig:StructureKnowledge}
\vspace{-3mm}
\end{figure}
\begin{example} \label{exp:2} 
As shown in Figure~\ref{fig:StructureKnowledge}, consider $\mathcal{G}_d = (\mathcal{V}_s \cup \mathcal{V}_b \cup \mathcal{V}_f,\; \mathcal{E}_{sb} \cup \mathcal{E}_{bf})$ with static attribute nodes $\mathcal{V}_s = \mathcal{V}_{\kappa} \cup \mathcal{V}_{\eta}$, behavior pattern nodes $\mathcal{V}_b = \{v_{b,1}, v_{b,2}, v_{b,3}\}$, and imputation function nodes $\mathcal{V}_f = \{v_{f,1}, v_{f,2}, v_{f,3}\}$.  
The pattern $v_{b,3} = (p^s, p^{\theta}, p^{\psi}, p^i, p^{\tau})$ corresponds to stable speed ($p^s$), gradual course change ($p^{\theta}$), low heading fluctuation ($p^{\psi}$), lane-following intent ($p^i$), and duration 1300--1350s ($p^{\tau}$).  
The edge $(v_{\eta,3}, v_{b,3}, 6) \in \mathcal{E}_{sb}$ (red edge in Figure~\ref{fig:StructureKnowledge}) shows that vessels with status ``under way using engine'' co-occurred with this pattern six times, while $(v_{b,3}, v_{f,1}, 6) \in \mathcal{E}_{bf}$ (blue edge in Figure~\ref{fig:StructureKnowledge}) indicates that the Linear Interpolation function $v_{f,1} = ( f, d(f) )$ successfully imputed this pattern six times.  
\end{example}

The SD-KG design is driven by two goals: making the knowledge both human-interpretable and LLM-usable, and ensuring that it supports accurate and efficient imputation.

\noindent \textbf{Human interpretability and LLM usability.} Semantically meaningful node types---static attributes, behavior patterns, and imputation functions---are readily understood by domain experts and simultaneously provide structured context that grounds LLM reasoning. The decomposition of behaviors into speed, course, heading, intent, and duration enhances transparency, and pairing imputation functions with descriptive semantics enables both humans and LLMs to reason about function--pattern appropriateness.

\noindent \textbf{Imputation accuracy and efficiency.} Imputation functions in $\mathcal{V}_f$ are derived from real AIS data, capturing empirically validated and executable strategies, while edge weights quantify statistical associations to guide function selection with data-grounded priors. Efficiency is achieved by compressing massive AIS data into a compact graph, enabling lightweight subgraph retrieval and direct function execution with incremental updates.
Further details on the construction and maintenance of SD-KG are provided in Section~\ref{sec:detailKnowledgeBase}. 

\begin{figure}[!tbp]
\centering
\includegraphics[width=1\linewidth]{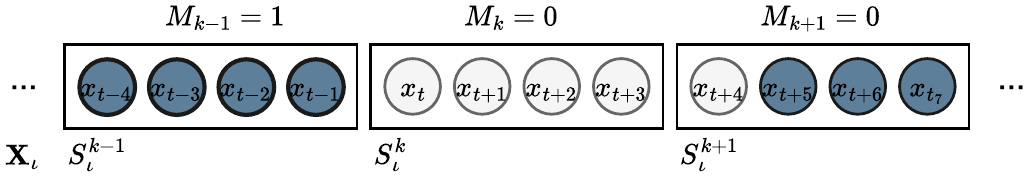}
\caption{Partition of an AIS record sequence $\mathbf{X}_{\iota}$ into minimal segments (4 records per segment).}
\label{fig:incompleteAISData}
\end{figure}

\subsection{Problem Definition}\label{sec:problemdefinition}
We formalize the problem in three parts. With AIS records defined (Definitions~\ref{def-1}--\ref{def-2}), we first capture data incompleteness via \emph{minimal segments} and an \emph{observation mask}. Next, for complete AIS data, each segment is distilled into a \emph{knowledge unit} and used to update the SD-KG $\mathcal{G}_d$. Finally, for incomplete segments, we define \emph{knowledge-driven trajectory imputation} that queries $\mathcal{G}_d$ under contextual constraints to retrieve candidate behavior patterns and imputation functions, selects and applies the most plausible function to reconstruct the segment, and returns the imputed segment together with its repair provenance.

\subsubsection{Incomplete AIS Data}
Missing data in AIS trajectories typically occurs as contiguous block-level gaps~\cite{lubkowski_assessment_2017}, where consecutive records are absent rather than isolated points. To capture this phenomenon in a structured way, we introduce the concept of a minimal segment, which serves as the atomic block for modeling both observed and missing intervals. 
\begin{definition}[\textbf{Minimal Segment}] \label{def-3}
For a vessel-specific AIS record sequence $\mathbf{X}_{\iota} = \langle x_1, x_2, \dots, x_T\rangle$,  
a minimal segment $\mathbf{S}_{[t, t+m-1]}$ is defined as a consecutive block of AIS records:
\begin{equation}
\mathbf{S}_{[t, t+m-1]} = \langle x_t, x_{t+1}, \dots, x_{t+m-1} \rangle,
\end{equation}
where $m$ is the fixed segment length, corresponding to the minimal unit for modeling missing trajectory data (default $m=20$).
\end{definition}

An AIS record sequence $\mathbf{X}_{\iota} = \langle x_1, \dots, x_T \rangle$  
can be partitioned uniquely into a concatenation of disjoint minimal segments:
\begin{equation}
\mathbf{X}_{\iota} = \mathbf{S}_{\iota}^1 \,\Vert\, \mathbf{S}_{\iota}^2 \,\Vert\, \dots \,\Vert\, \mathbf{S}_{\iota}^K,
\end{equation}
where each $\mathbf{S}_{\iota}^k = \mathbf{S}_{[t_k,\, t_k+m-1]}$ is the $k$-th minimal segment of fixed length $m$, ensuring that all segments are non-overlapping and together cover the entire AIS sequence of vessel $\iota$.

\begin{example}\label{exp:incompleteAISData-1}
Consider Figure~\ref{fig:incompleteAISData}, where the fixed segment length is $m=4$. Around index $t$, the AIS record sequence $\mathbf{X}_{\iota}$ is partitioned into three consecutive minimal segments: $\mathbf{S}_{\iota}^{k-1}=\langle x_{t-4},x_{t-3},x_{t-2},x_{t-1}\rangle$, $\mathbf{S}_{\iota}^{k}=\langle x_{t},x_{t+1},x_{t+2},x_{t+3}\rangle $, and $\mathbf{S}_{\iota}^{k+1}= \langle x_{t+4},x_{t+5},x_{t+6},x_{t+7}\rangle$.
\end{example}

\begin{definition}[\textbf{Observation Mask}] \label{def-segmentmask}
Given the fixed-length partition $\mathbf{X}_{\iota}=\mathbf{S}_{\iota}^1 \,\Vert\, \mathbf{S}_{\iota}^2 \,\Vert\, \dots \,\Vert\, \mathbf{S}_{\iota}^K$, the observation mask is the binary sequence
\begin{equation}
\mathbf{M}_{\iota} = \langle M_1,\dots,M_K \rangle ,
\end{equation}
where
\begin{equation}
M_k=
\begin{cases}
1 & \text{if } \forall\, j\in[t_k,t_k+m-1],~x_j \text{ has all attributes present}\\
0 & \text{otherwise}
\end{cases}
\end{equation}
\end{definition}

\begin{example}
Continuing Example~\ref{exp:incompleteAISData-1}, using the convention $M_k=0$ for a segment with missing records and $M_k=1$ for a fully observed segment, the configuration in Figure~\ref{fig:incompleteAISData} yields $M_{k-1}=1$, $M_{k}=0$, and $M_{k+1}=0$, i.e., the mask sequence $\mathbf{M}_{\iota}$ is $\langle \ldots,1,0,0,\ldots\rangle$.
\end{example}

\subsubsection{Updating of Structured Data-derived Knowledge}
A minimal segment $\mathbf{S}_{\iota}^k$ is a time-indexed sample and cannot directly update the SD-KG $\mathcal{G}_d$, which is organized over semantic node types and their relations. To bridge segment-level data and the graph-level representation, we introduce the \emph{knowledge unit} as the minimal semantic unit distilled from a segment: it includes a static attribute, a behavior pattern, and an imputation function, and it serves as the atomic granularity for SD-KG updating. 

\begin{definition}[\textbf{Knowledge Unit}] \label{def-ku}
A \emph{knowledge unit} $u$ is defined as a triple $\;u \;= ( v_s, v_b, v_f ) \;= (
      ( \iota, \eta, \chi, d, \ell, \beta, \kappa, \sigma ),\;
      ( p^s, p^{\theta}, p^{\psi}, p^i, p^{\tau} ),\; \\
      ( f, d(f) ))$, 
where $v_s$ is a static-attribute vector,
$v_b$ is a behavior pattern, and
$v_f$ is an imputation method.
\end{definition}
 
On this basis, we formally define how $\mathcal{G}_d$ is incrementally updated from complete segments.  

\begin{definition}[\textbf{Updating of Structured Data-derived Knowledge Graph}] \label{def-update-sdk}
Given a minimal segment $\mathbf{S}_{\iota}^k$ with $M_{\iota}^k = 1$,  the updating of SD-KG refers to the following procedure:
\begin{equation}
\mathcal{G}_d \;\leftarrow\; \mathrm{Update}(\mathcal{G}_d, u_{\iota}^k),
\end{equation}
where $u_{\iota}^k$ is the knowledge unit distilled from $\mathbf{S}_{\iota}^k$.  
Formally, the update consists of two steps:
\begin{itemize}[itemsep=2pt,leftmargin=12pt]
    \item \textbf{Knowledge distillation:}  
    $\mathbf{S}_{\iota}^k \mapsto u_{\iota}^k=( v_s,v_b,v_f)$, which compresses a raw segment into a static attribute $v_s$, a behavior pattern $v_b$, and an imputation method $v_f$.  
    \item \textbf{Graph integration and maintenance:}  The node set $\mathcal{V}_d$ and edge set $\mathcal{E}_d$ of $\;\mathcal{G}_d$ are updated by inserting or merging $v_s$, $v_b$, and $v_f$ and by adjusting the corresponding edges $(v_s,v_b)\in\mathcal{E}_{sb}$ and $(v_b,v_f)\in\mathcal{E}_{bf}$. Redundant nodes are merged, and edge weights are aggregated to ensure that $\mathcal{G}_d$ remains compact and efficient for retrieval.
\end{itemize}
\end{definition}

\begin{figure}[!tbp]
\centering
\includegraphics[width=0.9\linewidth]{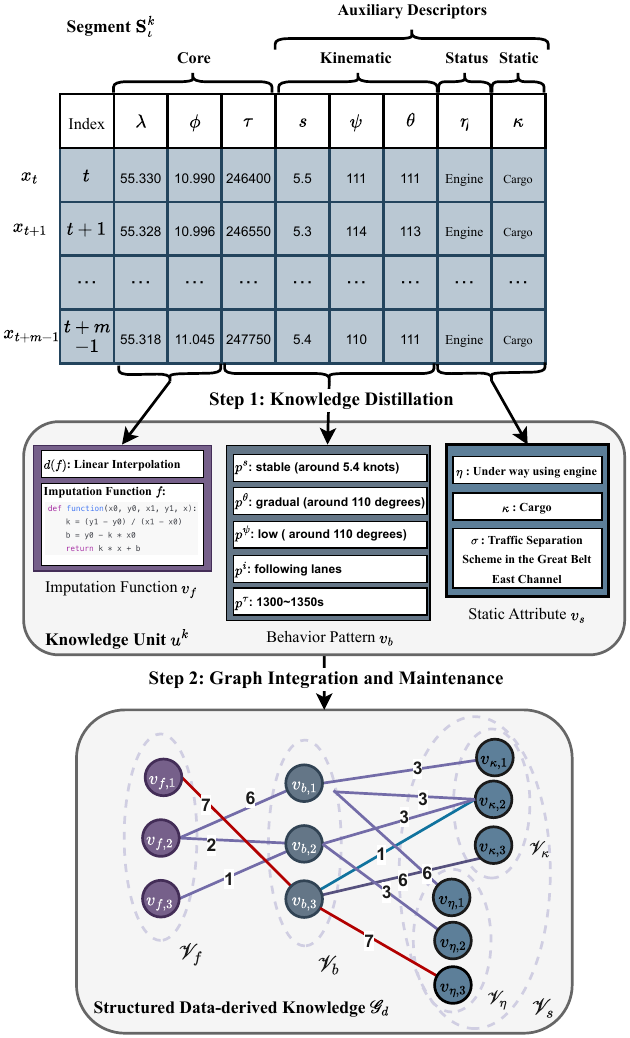}
\caption{Example of updating the SD-KG.}
\vspace{-3mm}
\label{fig:kb_construction}
\end{figure}
\begin{example}\label{exp:update-sdk}

In Figure~\ref{fig:kb_construction}, there is a segment $\mathbf{S}_{\iota}^k=\langle x_t,\ldots, x_{t+m-1}\rangle$ with $M_{\iota}^k=1$.  Within this segment, the speed $s$ remains stable at approximately $5.4$~kn, the course $\theta$ varies gradually around $110^\circ$, the heading $\psi$ exhibits mild fluctuations near $110^\circ$,  and both the navigation status and vessel type are constant, recorded as ``under way using engine'' and ``cargo''.

\noindent \textbf{Step 1: Knowledge distillation.}  
From the static and status-related attributes, together with the vessel’s geographic position, we derive the static attributes $v_s=( \iota, \eta, \kappa, \sigma )$ (listing only the attributes relevant to this example for brevity), describing a cargo vessel $\iota$ operating ``under way using engine'' near the Traffic Separation Scheme in the Great Belt East Channel.  
From the kinematic attributes, we extract the behavior pattern $v_b=( p^s, p^{\theta}, p^{\psi}, p^i, p^{\tau} )$, corresponding to stable speed, gradual course change, mild heading fluctuation, lane-following intent, and duration $1300$--$1350$s. To avoid redundancy in $\mathcal{V}_b$, the states $p^s,p^{\theta},p^{\psi},p^i,p^{\tau}$ are preferentially aligned with existing entries, in this case matching $v_{b,3} \in \mathcal{V}_b$ (continuing Example~\ref{exp:2}).  
Finally, based on the trajectory trend, we generate the imputation method $v_f=( f, d(f))$, where $f$ is the executable function that simulates the observed dynamics and $d(f)$ records its description.  

\noindent \textbf{Step 2: Graph integration and maintenance.}  Continuing Example~\ref{exp:2}, $v_b$ coincides with $v_{b,3}$, and the imputation function $v_f$ is semantically equivalent to $v_{f,1}$. The graph is updated accordingly: the weights of edges $(v_{\eta,3}, v_{b,3})$ and $(v_{b,3}, v_{f,1})$ are each incremented by $1$ (from $6$ to $7$, red edges), while a new edge $(v_{\kappa,2}, v_{b,3})$ is added with weight $1$ (blue edge).  
\end{example}
Further details on SD-KG updates are presented in Section~\ref{sec:detailKnowledgeBase}.

\subsubsection{Knowledge-Driven Trajectory Imputation}
Before specifying the imputation procedure, we formalize the target output. The outcome should not only reconstruct the missing trajectory but also produce structured \emph{repair provenance} that documents the reasoning behind each repair decision.

\begin{definition}[\textbf{Imputation Outcome with Repair Provenance}] \label{def-result-uk}
For a minimal segment $\mathbf{S}_{\iota}^k$ with $M_{\iota}^k=0$, the imputation outcome is defined as
$R_{\iota}^k = (\widehat{\mathbf{S}}_{\iota}^k,\; \mathcal{J}_{\iota}^k)$,
where $\widehat{\mathbf{S}}_{\iota}^k$ is the imputed trajectory segment and $\mathcal{J}_{\iota}^k$ is the accompanying repair provenance. The repair provenance $\mathcal{J}_{\iota}^k$ consists of:
\begin{itemize}[itemsep=2pt,leftmargin=12pt]
    \item \textbf{What} --- behavior-pattern estimation and rationale $(v_b^{*},\, \mathcal{J}_{\iota}^{k,b})$:
    $v_b^{*}$ is the estimated behavior pattern, and $\mathcal{J}_{\iota}^{k,b}$ provides contextual and statistical evidence from $\mathcal{G}_d$ supporting this estimation.

    \item \textbf{How} --- function selection and rationale $(v_f^{*},\, \mathcal{J}_{\iota}^{k,f})$:
    $v_f^{*}$ is the imputation function applied to obtain $\widehat{\mathbf{S}}_{\iota}^k$, and $\mathcal{J}_{\iota}^{k,f}$ contains statistical evidence from $\mathcal{G}_d$ justifying the choice of this function.

    \item \textbf{Why} --- causal explanation $\mathcal{J}_{\iota}^{k,h}$:
    a natural-language explanation of why the vessel exhibits the estimated behavior, grounded in the SDK context and generated by an LLM. 
\end{itemize}
\end{definition}

\begin{example}
Continuing Example~\ref{exp:1}, the imputed segment $\widehat{\mathbf{S}}_{\iota}^k$ is reconstructed as a smooth curve bridging the missing interval (upper-left in Figure~\ref{fig:intro-fi-1}). The repair provenance $\mathcal{J}_{\iota}^k$ records three components: \textbf{What}---the estimated behavior pattern $v_b^{*}$ is a decelerate-then-align maneuver with low heading fluctuation and queue-following intent, supported by evidence $\mathcal{J}_{\iota}^{k,b}$ showing this pattern occurs in about $68\%$ of comparable gaps. \textbf{How}---the chosen imputation function $v_f^{*}$ is ``CTR+SD'', with $\mathcal{J}_{\iota}^{k,f}$ recording that over $70\%$ of such gaps are best reconstructed by this function. \textbf{Why}---$\mathcal{J}_{\iota}^{k,h}$ explains that the maneuver arises from port-entry procedures, where vessels reduce speed and adjust heading to merge into inbound lanes under traffic separation rules. 
\end{example}

\begin{definition}[\textbf{Knowledge-Driven Trajectory Imputation}] \label{def-knowledge-imputation}
Given a minimal segment $\mathbf{S}_{\iota}^k$ with $M_{\iota}^k=0$ and contextual knowledge units $( u_{\iota}^{-}, u_{\iota}^{+} )$, knowledge-driven trajectory imputation produces an imputation outcome with repair provenance $R_i^k$:
\begin{equation}
R_{\iota}^k \;\leftarrow\; \mathrm{Impute}(\mathbf{S}_{\iota}^k, (u_{\iota}^{-}, u_{\iota}^{+}),  \mathcal{G}_d),
\end{equation}
where $u_{\iota}^{-}$ and $u_{\iota}^{+}$ are the contextual knowledge units extracted from the nearest complete segments preceding and succeeding $\mathbf{S}_{\iota}^k$, respectively. The procedure consists of:

\begin{itemize}[itemsep=2pt,leftmargin=12pt]
\item \textbf{Behavior pattern selection and rationale:}
Using the contextual knowledge units $( u_{\iota}^{-}, u_{\iota}^{+} )$ as query constraints, $\mathcal{G}_d$ is searched to retrieve a candidate set $\mathcal{C}_b \subseteq \mathcal{V}_b$.  A reasoning process then selects the most plausible behavior pattern $v_b^* \in \mathcal{C}_b$, and the retrieved set with associated evidence is recorded as $\mathcal{J}_{\iota}^{k,b}$.

\item \textbf{Imputation method selection, execution and rationale:}
Using the selected behavior pattern $v_b^*$ as query constraints, $\mathcal{G}_d$ is queried to retrieve $\mathcal{C}_f \subseteq \mathcal{V}_f$. An imputation procedure selects the most suitable method $v_f^* \in \mathcal{C}_f$ and applies it to obtain $\widehat{\mathbf{S}}_{\iota}^k$. The retrieved set and supporting statistics are recorded as $\mathcal{J}_{\iota}^{k,f}$.

\item \textbf{Causal explanation:}
Given the selected behavior pattern and imputation function $(v_b^*, v_f^*)$, along with the induced subgraph $\mathcal{G}_d[\mathcal{V}_s(\iota) \cup \{v_b^*\} \cup \{v_f^*\}]$ (where $\mathcal{V}_s(\iota)$ denotes the static attribute nodes of vessel $\iota$),
an LLM generates the natural-language explanation $\mathcal{J}_{\iota}^{k,h}$, providing causal reasoning for the reconstructed behavior grounded in SDK context.

\end{itemize}
\end{definition}

Further details on knowledge-driven trajectory imputation are provided in Section~\ref{sec:InterpretableImputation}.
\begin{figure}[!tbp]
\centering
\includegraphics[width=1.02\linewidth]{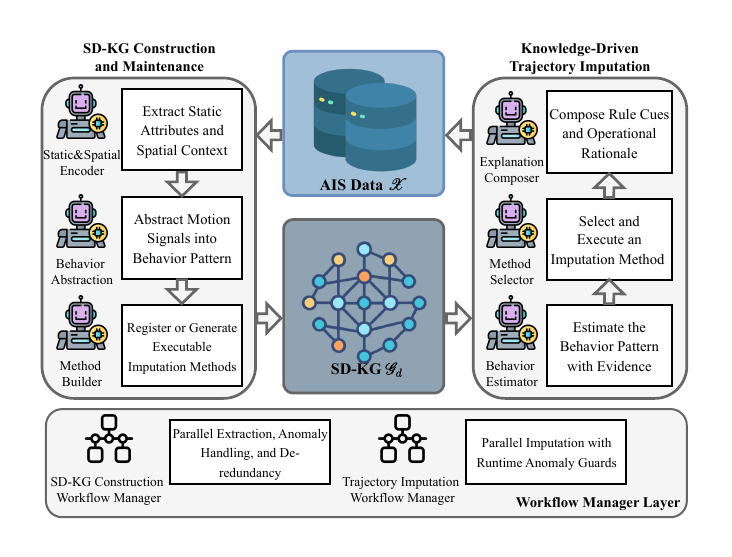}
\vspace{-4mm}
\caption{Overview of \textsf{VISTA}.}
\label{fig:VISTAFramework}
\vspace{-4mm}
\end{figure}
%----------------------------------------
\section{Knowledge-Driven Interpretable Vessel Trajectory Imputation} 
\label{sec:methodology}
\subsection{Overview of \textsf{VISTA}}
As illustrated in Figure~\ref{fig:VISTAFramework}, \textsf{VISTA} establishes a closed-loop that transforms raw AIS data into structured maritime knowledge and reuses this knowledge to reconstruct missing trajectories with interpretable reasoning. To ensure scalable and reliable execution across large-scale AIS datasets, \textsf{VISTA} is supported by a workflow management layer that enables parallel processing, anomaly handling, and redundancy control. 

\noindent\textbf{Data-Knowledge-Data Loop.} The core loop of \textsf{VISTA} consists of two interconnected stages: \emph{SD-KG Construction and Maintenance} and \emph{Knowledge-Driven Trajectory Imputation}. The first stage converts raw AIS data into structured knowledge through static and spatial encoding, behavior abstraction, and imputation method generation.
Each resulting knowledge unit, consisting of vessel attributes, behavior patterns, and validated imputation functions, is integrated into the SD-KG via node alignment and edge-weight updates, gradually building a compact and updatable knowledge repository.
The second stage reuses this knowledge to reconstruct incomplete trajectories. It estimates behavior patterns with graph-supported evidence, selects and executes suitable imputation functions, and composes explanations that link reconstructed behaviors to maritime rules and operational rationales.

\noindent\textbf{Workflow Manager Layer.} To handle the computational demands of large-scale AIS data and LLM-based inference, \textsf{VISTA} incorporates a dedicated workflow management layer operating in parallel with both stages. It includes two specialized managers: the \emph{SD-KG Construction Workflow Manager}, which orchestrates parallel knowledge extraction, anomaly handling, and de-redundancy, and the \emph{Trajectory Imputation Workflow Manager}, which manages concurrent imputation and runtime anomaly guard. 

\subsection{SD-KG Construction and Maintenance}
\label{sec:detailKnowledgeBase}
We construct and maintain the SD-KG through three components followed by a graph update. For each complete minimal segment, the \emph{Static and Spatial Encoder} produces static attributes and spatial context, the \emph{Behavior Abstraction} maps motion signals to a behavior pattern, and the \emph{Imputation Method Builder} registers or generates a validated executable method with a concise description. The outputs form a \emph{knowledge unit} $u=(v_s, v_b, v_f)$ as defined in Definition~\ref{def-ku}. Each knowledge unit is then integrated into the SD-KG through node alignment and edge-weight updates. 

\subsubsection{Static and Spatial Encoder}
The encoder organizes the static attribute $v_s=(\iota,\eta,\chi,d,\ell,\beta,\kappa,\sigma)$ into three classes and applies targeted processing. Discrete fields $\{\iota,\eta,\chi,\kappa\}$ are standardized, continuous fields $\{d,\ell,\beta\}$ are discretized into intervals, and spatial context $\{\sigma\}$ is inferred from geospatial surroundings. This design keeps the $\mathcal{V}_s$ in SD-KG compact, interpretable, and easy to maintain.

\noindent \textbf{Discrete Attributes.}
Vessel identifier $\iota$, navigation status $\eta$, hazardous cargo type $\chi$, and vessel type $\kappa$ are directly obtained from the corresponding static attributes of AIS records $x$ within the minimal segment $\mathbf{S}_{\iota}^k$ and standardized to canonical vocabularies. 

\noindent \textbf{Continuous Attributes.} Draught $d$, length $\ell$, and width $\beta$ are discretized into interval-based categories to form finite vocabularies $\mathcal{V}_d$, $\mathcal{V}_{\ell}$, and $\mathcal{V}_{\beta}$ and preserve interpretability. Specifically, $d$ is discretized into 2m intervals up to 12m, $\ell$ into 50m intervals up to 300m, and $\beta$ into 5m intervals up to 30m, each with an open-ended bin beyond the upper bound. These cutoffs reflect common maritime conventions~\cite{panama_vessel_2022,clarksons_bulk_2025} (e.g., Panamax thresholds at $\ell \approx 300$m, $\beta \approx 32$m, and $d \approx 12$m) while keeping sufficient \mbox{support}. 

\noindent \textbf{Spatial Context.}
The surrounding maritime environment is inferred by querying external sources based on coordinates $(\lambda,\phi)$ in AIS records within the minimal segment $\mathbf{S}_{\iota}^k$ to detect nearby infrastructures such as traffic separation schemes, shipping lanes, or anchorages (e.g., Overpass API~\cite{overpass_overpass_2025}). The most frequent category within $\mathbf{S}_{\iota}^k$ is selected as the representative $\sigma$.

% Since draught $d$ and vessel geometry (length $\ell$ and width $\beta$) are continuous attributes, we transform them into interval-based categories. This discretization yields finite vocabularies for $\mathcal{V}_d$, $\mathcal{V}_{\ell}$ and $\mathcal{V}_{\beta}$, ensuring that the SD-KG remains compact and semantically interpretable. Specifically, draught is divided into 2-meter intervals up to 16 m and one open-ended bin beyond, length is divided into 50-meter intervals up to 300 m and one open-ended bin beyond, and width is divided into 5–10 meter intervals up to 40 m and one open-ended bin beyond. 
% These cutoffs reflect common maritime conventions~\cite{panama_vessel_2022,clarksons_bulk_2025} (e.g., Panamax thresholds at $\ell \approx 300$ m, $\beta \approx 32$ m, and $d \approx 12$ m, with post-Panamax vessels exceeding one or more of these limits) while maintaining sufficient statistical support for each category.
% % To further ensure robustness, adjacent intervals are dynamically merged when data sparsity occurs, preventing over-fragmentation of nodes.

% For spatial context $\sigma$, we determine the surrounding maritime environment by querying external sources (e.g., Overpass API~\cite{overpass_overpass_2025}) based on the coordinates $(\lambda, \phi)$ in $x$, which detect nearby infrastructures such as traffic separation schemes, shipping lanes, or anchorage areas. The spatial category most frequently occurring within $\mathbf{S}_{\iota}^k$ is then selected as the representative spatial context $\sigma$.

\subsubsection{Behavior Abstraction} \label{sec:ku-extraction}
This component converts motion signals within the minimal segment into a discrete, human-interpretable behavior pattern $v_b=(p^s,p^{\theta},p^{\psi},p^{i},p^{\tau})$ and constrains the vocabulary growth for interpretability and efficient retrieval.

\noindent \textbf{Kinematic Tokens $(p^s,p^{\theta},p^{\psi})$.}
Speed $s$, course over ground $\theta$, and heading $\psi$ exhibit diverse dynamics; direct tokenization yields opaque categories and uncontrolled growth. An LLM-assisted procedure analyzes the temporal evolution in $\mathbf{S}_{\iota}^k$ and maps them to discrete variables $(p^s,p^{\theta},p^{\psi})$. The vocabulary size is controlled by finite sets $\mathcal{P}^s$, $\mathcal{P}^{\theta}$, and $\mathcal{P}^{\psi}$ with a deduplication rule that reuses semantically equivalent tokens before adding new ones.

\noindent \textbf{Intent Token ($p^i$).}
Navigation intent $p^i$ is inferred from static attributes $v_s$ and contextual cues $(p^s,p^{\theta},p^{\psi},p^{\tau})$ through LLM analysis. We maintain a finite intent set $\mathcal{P}^i$ with the same deduplication rule; ambiguous cases fall back to an undetermined intent.

\noindent \textbf{Duration Token ($p^{\tau}$).}
Duration token $p^{\tau}$ is computed from the segment time span and discretized into fixed-width 50s intervals with an open-ended bin for long gaps (i.e., $[3600, +\infty)$). 
% \warning{Implementation details, prompt templates of LLM, and token deduplication rules appear in Appendix~\ref{prompt-BehaviorPatternsExtraction}.}

\subsubsection{Imputation Method Builder} \label{sec:methodbuilder}
The Imputation Method Builder produces the imputation method $v_f=(f,d(f))$, where $f$ is an executable function and $d(f)$ is a concise usage description. 
% It retrieves existing methods to avoid duplication, validates candidates by execution, and finalizes accepted methods with a short description aligned to the behavior pattern. 
Given a minimal segment $\mathbf{S}_{\iota}^k$, $v_f$ is generated through a three-step process.

\noindent \textbf{Function Proposal.} Based on $\mathbf{S}_{\iota}^k$ and the associated $(v_s,v_b)$, 
we first retrieve the most relevant imputation method $v_f \in \mathcal{V}_f$ using the retrieval scheme described in Section~\ref{sec:MethodSelector}, to prevent unnecessary duplication and keep $\mathcal{V}_f$ compact. If no suitable candidate is found, an LLM-based function generator produces a new executable function $f$ that directly fits the observed trajectory.

\noindent \textbf{Execution and Validation.} The proposed function $f$ is executed on $\mathbf{S}_{\iota}^k$. At each timestamp we compute the absolute errors in latitude and longitude, then average them to obtain the overall fitting error $e(f) = \tfrac{1}{2}(\mathrm{MAE}_{\phi}+\mathrm{MAE}_{\lambda})$. The function is accepted if $e(f)$ does not exceed a predefined threshold of $\varrho_f = 3e-3$. If the error exceeds this threshold, the LLM-based function generator is provided with diagnostic feedback (error summaries) and asked to refine its proposal. This is repeated at most $\varrho_r$ times (default 3).  

\noindent \textbf{Description and Alignment.} For each accepted function, if no descriptive element $d(f)$ is available, the LLM-based function explainer generates $d(f)$ conditioned on the executable code $f$ and the selected behavior pattern $v_b$. The description specifies the intended use, key assumptions, and parameter definitions, thereby improving the interpretability of the imputation method. 
% \warning{For implementation details, including the LLM prompt templates and the rules for generating, validating, and refining imputation functions, see Appendix~\ref{prompt-ImputationFunctionGeneration}.}

\subsubsection{SD-KG Integration and Maintenance} \label{sec:sdkg-build}
Once knowledge units are extracted, the SD-KG is incrementally updated through two operations: node integration and edge maintenance.  

\noindent \textbf{Node Integration.}  
Since static attributes, behavior patterns, and imputation methods have already undergone de-redundancy during knowledge unit extraction, the integration process is straightforward. 
For each static attribute, if a newly generated token does not yet exist, a corresponding node is created and added to $\mathcal{V}_s$. 
Similarly, if the behavior pattern $v_b$ is not present in $\mathcal{V}_b$, or the imputation method $v_f$ is not present in $\mathcal{V}_f$, a new node is inserted. 

\noindent \textbf{Edge Maintenance.}  
For each knowledge unit $u=(v_s,v_b,v_f)$, edges $(v_s,v_b)$ and $(v_b,v_f)$ are updated. 
If the edge does not exist, it is created with an initial weight of $1$. 
If it already exists, the weight is incremented by $1$, capturing the frequency of observed co-occurrences. 
This ensures that the SD-KG preserves both structural completeness and statistically grounded associations while remaining compact and retrievable.  

Details of the LLM prompts and the overall SD-KG construction and maintenance procedure are provided in Appendix~\ref{app:M1-M3}.

\subsection{Knowledge-Driven Trajectory Imputation}
\label{sec:InterpretableImputation}
% With the SD-KG $\mathcal{G}_d$ in place, we formalize knowledge driven trajectory imputation for an incomplete minimal segment $\mathbf{S}_{\iota}^k$ in more detail. As defined in Definition~\ref{def-knowledge-imputation}, the procedure has three parts: behavior pattern selection, imputation method selection and implementation, and underlying cause generation for a human friendly explanation. In what follows, we show how contextual constraints query the SD-KG to retrieve and rank candidate patterns, how the selected method reconstructs the gap and is validated, and how the explanation is derived from the induced subgraph.
With the SD-KG $\mathcal{G}_d$ in place, we formalize knowledge-driven trajectory imputation for an incomplete minimal segment $\mathbf{S}_{\iota}^k$. The procedure comprises three components: \emph{Behavior Estimator}, which uses contextual constraints to query $\mathcal{G}_d$, rank candidates, and select the behavior pattern with supporting evidence; \emph{Method Selector}, which retrieves an imputation method aligned with the chosen pattern, and executes it to reconstruct the gap; and \emph{Explanation Composer}, which derives a human-friendly explanation by combining a regulatory rule cue with an operational protocol rationale from the induced subgraph of SD-KG and contextual information. In what follows, we detail each component.

\subsubsection{Behavior Estimator} \label{sec:behaviorEstimator}
The Behavior Estimator operates in two steps: \emph{Candidate Set Generation}, which queries the SD-KG to produce a ranked shortlist of plausible behavior patterns, and \emph{LLM-driven Selection and Rationale}, which selects the final pattern and articulates a supporting rationale.

\noindent \textbf{Candidate Set Generation.} 
Given an incomplete segment $\mathbf{S}_{\iota}^k$ with context $( u_{\iota}^{-},u_{\iota}^{+} )$, it infers the vessel-specific static attribute set $\mathcal{V}^{k}_s(\iota)$ from $(u_{\iota}^{-},u_{\iota}^{+})$, since static attributes vary slowly over time. It then queries the SD-KG to retrieve behavior candidates via static–behavior edges $C_b$:
\begin{equation}\label{eq:Cb}
\mathcal{C}_b \;=\; \{\, v_b\in\mathcal{V}_b \ |\ \sum_{v\in \mathcal{V}^{k}_s(\iota)} w_{sb}(v, v_b) > 0 \,\},
\end{equation}
and it assigns each candidate a graph prior based on normalized log-additive support:
\begin{equation} \label{eq:nodePrior}
\pi(v_b) \;=\;
\frac{\displaystyle\sum_{v\in \mathcal{V}^{k}_s(\iota)}\!\log\!\big(w_{sb}(v,v_b)+1\big)}
{\displaystyle\sum_{v_b'\in \mathcal{C}_b}\;\sum_{v\in \mathcal{V}^{k}_s(\iota)}\!\log\!\big(w_{sb}(v,v_b')+1\big)},
\end{equation}
% \begin{equation} \label{eq:nodePrior}
% \pi(v_b) \;=\; (\sum_{v\in \mathcal{V}^{k}_s(\iota)} w_{sb}(v, v_b))) * w_{sb}(\sigma, v_b)
% \end{equation}
where $w_{sb}$ is the weight of edge $(v,v_b,w)$. It keeps the top-$K$ candidates by $\pi(v_b)$ as the shortlist $\mathcal{C}_b^{K}$ (default $K{=}5$).

% \begin{example} Let $\mathcal{V}^{k}_s(\iota)=\{\text{``cargo''},\,\text{``under way using engine''}\}$ and consider two candidates $v_{b,1}$ and $v_{b,2}$. Suppose $w_{sb}(\text{``cargo''},v_{b,1})=20$ and $w_{sb}(\text{``under way using engine''},v_{b,1})=0$, while $w_{sb}(\text{``cargo''},v_{b,2}) \\ = 5$ and $w_{sb}(\text{``under way using engine''},v_{b,2})=5$. 
% By Equation~\ref{eq:nodePrior}, the smoothed multiplicative supports are $21$ for $v_{b,1}$ and $36$ for $v_{b,2}$. After normalization, $\pi(v_{b,2})>\pi(v_{b,1})$, indicating that balanced support across attributes is preferred over a single dominant cue.
% \end{example}

\noindent\textbf{LLM-driven Selection and Rationale.}
It first constructs the induced subgraph $\mathcal{G}_{\iota,b}^{K} = \mathcal{G}_d [\mathcal{V}^{k}_s(\iota) \cup \mathcal{C}_b^{K}]$, 
serializes it in a graph description language (i.e., DOT~\cite{wiki_DOT_2025}), and supplies it together with the contextual behavior patterns $v_b^{-}$ and $v_b^{+}$ extracted from $u_{\iota}^{-}$ and $u_{\iota}^{+}$. 
It then selects a final pattern $v_b^{*}\in\mathcal{C}_b^{K}$ based on the LLM's analysis of this information. 
% If no candidate is satisfactory, we propose an estimated pattern $\tilde v_b$ with tokens $(p^s,p^{\theta},p^{\psi},p^{i},p^{\tau})$, which we de-duplicate against $\mathcal{P}^s,\mathcal{P}^{\theta},\mathcal{P}^{\psi},\mathcal{P}^i$ following Section~\ref{sec:ku-extraction}.

Alongside the selection, it also obtains a two-part rationale: 
i) \emph{graph support}, which highlights the most informative edges in $\mathcal{G}_{\iota,b}^{K}$ and their weights that favor the choice; and 
ii) \emph{contextual justification}, which explains the consistency with $v_b^{-}$, $v_b^{+}$, and the gap’s boundary conditions. 
These two parts constitute $\mathcal{J}_{\iota}^{k,b}$. 
% \warning{Prompt templates appear in Appendix~\ref{prompt-BehaviorSelection}.}

\subsubsection{Method Selector} \label{sec:MethodSelector}
% This component parallels Behavior Estimator but omits contextual information during selection, because the chosen behavior $v_b^{*}$ already encodes the relevant kinematics and intent, and method quality is validated directly by execution, making extra context redundant.
This component operates similarly to the Behavior Estimator but omits contextual information during selection, as the selected behavior $v_b^{*}$ already encapsulates the relevant kinematic and intent information. 
Given the selected behavior $v_b^{*}$, it retrieves candidate functions through the behavior–function edges and scores them using the same prior construction as in Equations~\ref{eq:Cb} and \ref{eq:nodePrior}, yielding a ranked shortlist $\mathcal{C}_f^{K}$. It then builds the induced subgraph $\mathcal{G}_{\iota,f}^{K}=\mathcal{G}_d [\{v_b^*\} \cup \mathcal{C}_f^{K}]$, serializes it in DOT, and selects the best method $v_f^{*}\in\mathcal{C}_f^{K}$ by leveraging the LLM’s analysis of graph evidence and compatibility with the chosen behavior $v_b^{*}$. Next, it executes $f \in v_f^{*}$ to obtain the reconstruction $\widehat{\mathbf{S}}_{\iota}^k$. Alongside the selection, it records graph support by citing the informative edges within $\mathcal{G}_{\iota,f}^{K}$, which forms the rationale $\mathcal{J}_{\iota}^{k,f}$. 
% \warning{For the LLM prompt templates, see Appendix~\ref{prompt-MethodSelection}.}

\subsubsection{Explanation Composer} \label{sec:underlying-cause}
The Explanation Composer delivers a human-friendly explanation that helps downstream tasks internalize two kinds of domain knowledge: a \emph{Regulatory-rule cue} (under which rules), and an \emph{Operational-protocol rationale} (why it typically happens here). It analyzes a compact evidence view composed of the selected behavior $v_b^{*}$, the selected method $v_f^{*}$, static attributes $\mathcal{V}_s(\iota
)$, spatial context $\sigma$, boundary patterns $(v_b^{-}, v_b^{+})$, and the induced subgraph $\mathcal{G}_d[\mathcal{V}_s(\iota) \cup \{v_b^*\} \cup \{v_f^*\}]$. Specifically, the explanation consists of the following components:

\noindent \textbf{Regulatory-rule Cue.} A minimal yet sufficient indication of the governing navigation or traffic rule in context, expressed as a rule label with applicability conditions and a spatial anchor (e.g., ``TSS inbound-lane compliance; applies at port-entrance traffic separation; near pilot boarding line''). It is activated by the spatial context $\sigma$ (e.g., TSS, anchorage, pilot line, speed-restricted area), static vessel attributes (e.g., vessel type, vessel size), and the intent in $v_b^{*}$. The cue is generated in a concise, human-readable form, highlighting only the most relevant rules for the current scenario.

% \noindent \textbf{Operational-protocol Rationale.}
% A procedure level explanation of why the behavior pattern is typically expected here and how it relates to the rule cue. It emphasizes practical workflow logic such as sequencing, merging, pilot approach, collision avoidance, or weather avoidance. The LLM semantically aligns $v_b^{*}$ with the rule cue and the spatial context $\sigma$ to propose the most explanatory operational routine, and briefly rules out salient alternatives (e.g., why this is not a swerve for collision avoidance). The rationale does not conflate the imputation method with the operational reason; it remains independent of the chosen function and focuses on institutional and procedural logic.
\noindent \textbf{Operational-protocol Rationale.}  
A procedural-level explanation of why the selected behavior pattern typically occurs in this context and how it relates to the corresponding rule cue. It emphasizes operational logic such as sequencing, merging, pilot approach, collision avoidance, or weather avoidance. By semantically aligning $v_b^{*}$ with the rule cue and spatial context $\sigma$, the reasoning process infers the most plausible operational routine and briefly contrasts it with salient alternatives (e.g., explaining why this is not a swerve for collision avoidance). The rationale remains independent of the imputation method, focusing instead on institutional and procedural logic that underpins vessel operations.

Further details (i.e., prompts and overall procedure) are provided in Appendix~\ref{app:knowledgedriventrajectoryimputation}.

\begin{table*}[!htbp]
  \centering
  \caption{Overall effectiveness comparison on AIS-DK and AIS-US datasets.}
  \vspace{-4mm}
  \label{tab:main-1}
  \setlength{\tabcolsep}{4.2pt}
  \small
  \begin{tabular}{l|ccccc|ccccc}
    \toprule
    \multirow{2}{*}{Method} 
    & \multicolumn{5}{c|}{\textbf{AIS-DK}} 
    & \multicolumn{5}{c} {\textbf{AIS-US}} \\
    \cmidrule{2-11}
    & MAE(Lat) & RMSE(Lat) & MAE(Lon) & RMSE(Lon) & MHD 
    & MAE(Lat) & RMSE(Lat) & MAE(Lon) & RMSE(Lon) & MHD \\
    \midrule
    Lin-ITP              & 2.650e-3 & 6.386e-2 & 2.708e-3 & 5.379e-2 & 0.3978
                               & 1.497e-2 & 1.453e-1 & 3.336e-2 & 4.277e-1 & 3.9662\\
    Akima Spline   & 2.127e-3 & 5.984e-2 & 2.188e-3 & 4.833e-2 & 0.3195
                               & 1.274e-2 & 1.510e-1 & 3.070e-2 & 4.539e-1 & 3.5812 \\
    Kalman Filter  & 2.026e-3 & 5.664e-2 & 2.158e-3 & 4.799e-2 & 0.3085
                               & 1.250e-2 & 1.294e-1 & 2.742e-2 & 3.848e-1 & 3.2864 \\
    Multi-task AIS &  1.992e-3   & 5.364e-2  & 2.086e-3  &    4.692e-2  & 0.2932                   
                               &  1.113e-2   & 1.164e-1  & 2.593e-2  &    3.684e-1  & 3.1316    \\
    MH-GIN            &  \underline{1.906e-3}   & \underline{4.923e-2}  & \underline{1.895e-3}  &    \underline{4.417e-2}  & \underline{0.2836} 
                               &  \underline{8.728e-3}   & \underline{8.765e-2}  & \underline{2.231e-2}  &    \underline{3.292e-1}  & \underline{2.2164}    \\
    KAMEL             & 1.975e-3    & 5.123e-2  & 1.950e-3  &    4.539e-2  & 0.2875
                               &  9.875e-3   & 9.123e-2  & 2.450e-2  &    3.439e-1  & 2.6268 \\
    Qwen-plus-th  & 5.208e-3 & 5.734e-2 & 8.694e-3 & 5.063e-2 & 0.9139
                               & 8.132e-2 & 2.599e+0 & 3.027e-2 & 3.751e-1 & 10.9758\\
    Qwen-flash-th & 2.161e-2 & 5.622e-1 & 1.402e-1 & 3.852e+0 & 9.5032
                               & 7.106e-2 & 8.314e-1 & 1.478e+1 & 6.091e+1 &  482.6334 \\
    GLM-4.5-th & 2.641e-3 & 4.929e-2 & 3.675e-3 & 4.895e-2 & 0.3278
                         & 9.896e-3 & 9.431e-2 & 2.464e-2 & 3.467e-1 & 2.9531 \\
    GLM-4.5-air-th &  6.265e-3 & 6.201e-2 & 8.738e-3 & 5.122e-2 & 1.0216
                                & 2.301e-2 & 1.579e-1 & 3.075e-2 & 3.970e-1 & 4.9803 \\
    \midrule
    \textsf{VISTA}      &  \textbf{1.817e-3} & \textbf{4.324e-2} & \textbf{1.123e-3} & \textbf{4.027e-2} & \textbf{0.2418}
                     & \textbf{7.587e-3} & \textbf{2.050e-2} & \textbf{1.027e-2}  & \textbf{2.942e-2} & \textbf{1.4532} \\
    \bottomrule
  \end{tabular}
  % \vspace{-3mm}
\end{table*}

\begin{table}[!tbp]
  \centering
  \small
  \caption{Overall time cost comparison.}
  % \vspace{-2mm}
  \label{tab:timeOnly}
  \setlength{\tabcolsep}{1.1 pt}
  \begin{tabular}{l|cccc|c}
    \toprule
    Dataset & Qwen-plus-th & Qwen-flash-th & GLM-4.5-th & GLM-4.5-air-th & \textsf{VISTA} \\
    \midrule
    \textbf{AIS-DK} & 30:15:11 & 14:14:04 & 91:07:30 & 25:07:30 & 6:32:37 \\
    \textbf{AIS-US} & 24:09:11 & 12:10:08 &  88:52:08 & 22:22:52 & 6:01:11 \\
    \bottomrule
  \end{tabular}
  % \vspace{-3mm}
\end{table}

% \begin{table*}[!htbp]
%   \centering
%   \caption{Computation Cost of Autoregressive LLMs on Denmark and US Datasets}
%   \setlength{\tabcolsep}{2.2pt}\small
%   \begin{tabular}{l|cc|cc}
%     \hline
%     \multirow{2}{*}{Model} 
%     & \multicolumn{2}{c|}{\textbf{Denmark Dataset}} 
%     & \multicolumn{2}{c}{\textbf{US Dataset}} \\
%     & Tokens & Time & Tokens & Time \\
%     \hline
%     Qwen-plus-Thinking & 1260280 & 39:54 & 1286000 & 44:32 \\
%     Qwen3-32B-Thinking & 1260485 & 34:34 & 1285993 & 38:11 \\
%     GLM-4-plus-Thinking & 960237 & 32:51 & 943860 & 35:07 \\
%     GLM-4-air-Thinking & 960900 & 40:01 & 945300 & 31:56 \\
%     \textsf{VISTA}       \\
%     % DeepSeek-v3 & 846720 & 34:05 & 864000 & 42:04 \\
%     \hline
%   \end{tabular}
%   \label{tab:llm_time_token}
% \end{table*}
\subsection{Workflow Management Layer}
\label{sec:scalable-parallel}
While the previous sections establish how \textsf{VISTA} extracts knowledge and performs trajectory imputation, both stages rely heavily on LLM inference---an operation that is inherently slow and occasionally unreliable due to stochastic failures (e.g., malformed outputs, null responses, or execution errors)~\cite{ahn2025prompt,tan2025too,agrawal2024taming}. To ensure scalability and robustness when processing large AIS data, we design a workflow management layer that automates task scheduling, anomaly handling, and de-redundancy. This layer includes two complementary components: the \emph{SD-KG Construction Workflow Manager} and the \emph{Trajectory Imputation Workflow Manager}.

\subsubsection{SD-KG Construction Workflow Manager} \label{sec:build-manager}
% The SD-KG Construction Workflow Manager coordinates a two–stage parallel workflow, \emph{extraction} and \emph{de-redundancy}. It comprises three modules: \emph{Stack-Based Scheduler}, \emph{Anomaly Guard}, and \emph{De-redundancy Processor}.
The SD-KG Construction Workflow Manager coordinates a two-stage parallel work\-flow---\emph{extraction} and \emph{de-redundancy}---through three modules: \emph{Stack-Based Scheduler}, \emph{Anomaly Guard}, and \emph{De-redundancy Processor}. 

\noindent \textbf{Stack-Based Scheduler.} 
All minimal segments are ordered by timestamp and pushed as job tuples onto a compute stack $\mathcal{S}_c$, while a second stack $\mathcal{S}_d$ buffers validated micro-batches before de-redundancy. At runtime, the scheduler pops up to $b$ jobs from $\mathcal{S}_c$ to form a micro-batch $\mathcal{B}_t$ and dispatches them to $|\mathcal{B}_t|$ worker threads. Each worker executes the extraction pipeline to produce a per-segment knowledge unit $u = (v_s, v_b, v_f)$, incrementing the retry counter $c$ on failure and discarding the job when $c > \varrho_c$. 
Validated outputs are packed as a micro-batch $\widehat{\mathcal{B}}_t$ and pushed onto the second stack $\mathcal{S}_d$, 
which acts as a decoupling buffer between extraction and de-redundancy. 
This preserves temporal order while isolating consolidation from extraction-rate fluctuations.

% All minimal segments are ordered by timestamp and pushed as job tuples onto a compute stack $\mathcal{S}_c$, while a second stack $\mathcal{S}_d$ buffers validated micro-batches before de-redundancy. At runtime, the scheduler pops up to $b$ jobs from $\mathcal{S}_c$ to form a micro-batch $\mathcal{B}_t$ and dispatches them to $|\mathcal{B}_t|$ worker threads. 
% Each worker executes the extraction pipeline to produce a per-segment knowledge unit $u_{\iota}^k = (v_s, v_b, v_f)$, incrementing the retry counter $c$ on failure and discarding the job when $c > \varrho_c$. 
% Validated outputs are packed as a micro-batch $\widehat{\mathcal{B}}_t$ and pushed onto the second stack $\mathcal{S}_d$. 
% The second stack acts as a decoupling buffer between extraction and de-redundancy, preserving temporal order and metadata while isolating \emph{de-redundancy} from \emph{extraction}. 
% This design allows the de-redundancy processor to operate asynchronously, merge semantically equivalent patterns across adjacent batches, and update the SD-KG in larger, more consistent units.

\noindent \textbf{Anomaly Guard.} 
The generated knowledge unit $u$ may contain format errors or invalid content due to response timeouts, hallucinated fields, schema drift, or incomplete outputs. 
To protect the SD-KG, each produced unit is validated before any graph update. 
The validator checks that (i) the output is non-empty, (ii) the schema and format are correct, and (iii) all $(v_s,v_b,v_f)$ components are present. 
Failed items are retried with bounded attempts: if $c<\varrho_c$, the job is re-queued with $c\!\leftarrow\!c{+}1$; otherwise, it is quarantined and excluded from subsequent stages. 
The validated subset of each batch is forwarded to the de-redundancy stack $\mathcal{S}_d$.

\noindent \textbf{De-redundancy Processor.} 
Before inserting new knowledge units into the SD-KG, redundancy elimination is applied to maintain compactness and prevent uncontrolled vocabulary growth. 
Validated batches are processed in parallel through: 
1) \emph{behavior-token canonicalization} merges semantically redundant tokens within behavior patterns using contextual and linguistic similarity analysis, ensuring that only canonical and representative tokens are retained across $\mathcal{P}^s$, $\mathcal{P}^{\theta}$, $\mathcal{P}^{\psi}$, and $\mathcal{P}^{i}$; 
2) \emph{function-equivalence testing} evaluates whether two imputation functions realize the same operational logic despite syntactic or parametric differences. 

Finally, validated and de-duplicated knowledge units are written to the SD-KG. New nodes ($v_s, v_b, v_f$) are inserted if absent, and edge weights $(v_s, v_b)$, $(v_b, v_f)$ are incremented atomically to maintain statistical coherence.

\subsubsection{Trajectory Imputation Workflow Manager}
\label{sec:impute-manager}
The Trajectory Imputation Workflow Manager mirrors the SD-KG Construction Workflow Manager but targets the reconstruction of incomplete minimal segments rather than graph construction. It omits the \emph{De-redundancy Processor}, as imputation does not produce new knowledge units requiring de-duplication.

\noindent\textbf{Stack-Based Scheduler.} 
All incomplete minimal segments are timestamp-sorted and pushed onto a stack $\mathcal{S}_{i}$ in descending order. The scheduler repeatedly pops a batch of size $b$ and executes the imputation pipeline in parallel. Each job runs the three components in Section~\ref{sec:InterpretableImputation} to reconstruct the missing segment using the SD-KG. A failure counter is maintained per job; when it exceeds the retry threshold $\varrho_i$, the job is aborted and logged for offline diagnosis.

\noindent\textbf{Anomaly Guard.} 
Imputation anomalies primarily arise from response timeouts or invalid model outputs. Each job therefore undergoes two basic checks: 1) the response is non-empty (i.e., no request timeout), and 2) the selected behavior pattern $v_b^*$ is correctly retrieved from the SD-KG and the imputation function $v_f^*$ is executable. Jobs failing either check are retried up to $\varrho_r$ times before being quarantined for offline inspection.

Finally, the selected imputation methods are executed to reconstruct the missing trajectory segments. The resulting trajectories, along with their explanations, are recorded for interpretability and traceability. Further details (e.g., prompt about redundancy and overall procedures) are provided in Appendix~\ref{app:workflowmanagementlayer}.

% These reconstructions with explanations serve as the final outputs of the imputation workflow and can be further utilized in downstream analytical or regulatory tasks.

% \clearpage
\begin{table*}[!tbp]
  \centering
  \small
  \caption{Performance of \textsf{VISTA} variants with different LLM configurations on AIS-DK and AIS-US datasets.}
  \label{tab:ablation-combined}
  % \vspace{-3mm}
  \setlength{\tabcolsep}{1.8pt}
  \begin{tabular}{ccc|ccccc|ccccc}
    \toprule
    \multirow{2}{*}{Analysis} & \multirow{2}{*}{Programming} & \multirow{2}{*}{Decision} 
    & \multicolumn{5}{c|}{\textbf{AIS-DK}} 
    & \multicolumn{5}{c}{\textbf{AIS-US}} \\
    \cmidrule(lr){4-8} \cmidrule(lr){9-13}
    & & & MAE(Lat) & RMSE(Lat) & MAE(Lon) & RMSE(Lon) & MHD 
      & MAE(Lat) & RMSE(Lat) & MAE(Lon) & RMSE(Lon) & MHD \\
    \midrule
    Qwen-plus   & Qwen-plus & Qwen-plus 
    & 1.817e-3 & 4.324e-2 & 1.123e-3  & 4.027e-2 & 0.2418
    & 7.587e-3 & 2.050e-2 & 1.027e-2  & 2.942e-2 & 1.4532 \\
    
    Qwen-plus   & Qwen-plus & Qwen-flash 
    & 5.706e-3   & 8.177e-2   & 5.578e-1   & 2.055e+1 & 2.8990
    & 2.352e-2   & 3.418e-1   & 5.104e+0   & 1.501e+1 & 17.4384 \\

    Qwen-plus   & Qwen-flash & Qwen-plus
    & 2.872e-3 & 3.248e-2 & 5.818e-3 & 6.126e-2 & 0.5255
    & 1.198e-2 & 4.637e-2 & 5.340e-2 & 8.913e-2 & 3.1527 \\

    Qwen-flash & Qwen-plus & Qwen-plus
    & 1.902e-3   & 5.826e-2   & 1.216e-3   & 4.718e-2   & 0.2672
    & 8.135e-3   & 2.263e-2  & 1.106e-2   & 3.354e-2   & 1.5847 \\

    \bottomrule
  \end{tabular}
  % \vspace{-2mm}
\end{table*}

\begin{table}[!tbp]
  \centering
  \small
  \caption{Ablation study of the workflow management layer.}
  \label{tab:ablation-managelayer}
  % \vspace{-2mm}
  \setlength{\tabcolsep}{2.3pt}
  \begin{tabular}{c|cc|cc|cc|cc}
    \toprule
    \multirow{3}{*}{\textbf{Batch Size}} 
      & \multicolumn{4}{c|}{\textbf{AIS-DK}} 
      & \multicolumn{4}{c}{\textbf{AIS-US}} \\
    \cmidrule(lr){2-5} \cmidrule(lr){6-9}
      & \multicolumn{2}{c|}{Vanilla} & \multicolumn{2}{c|}{w/o DR}
      & \multicolumn{2}{c|}{Vanilla} & \multicolumn{2}{c}{w/o DR} \\
    \cmidrule(lr){2-3} \cmidrule(lr){4-5} \cmidrule(lr){6-7} \cmidrule(lr){8-9}
      & \textbf{Time} & \textbf{Size} 
      & \textbf{Time} & \textbf{Size} 
      & \textbf{Time} & \textbf{Size} 
      & \textbf{Time} & \textbf{Size} \\
    \midrule
    8  & 13:10:21 & 1,016 & 10:50:22 & 1,248 & 12:06:39 & 1,882 & 10:07:29 &  3,406 \\
    16 & 6:32:37  & 1,105 & 5:27:39  & 1,400 & 6:01:11  & 2,057 & 5:10:26  & 4,109 \\
    \bottomrule
  \end{tabular}
  % \vspace{-3mm}
\end{table}

\section{Experimental Study}
\label{sec:experiment}
\subsection{Settings}
\subsubsection{Datasets}
We use two AIS datasets: AIS-DK from the Danish Maritime Authority~\cite{aisdkdataset} and AIS-US from the National Oceanic and Atmospheric Administration~\cite{aisusdataset}. AIS-DK covers Danish waters in March 2024, including major routes in the Baltic and North Seas, with 10,000 vessel sequences and 2,000,000 AIS records from 348 vessels, each lasting on average 0.5 hours. AIS-US covers U.S. coastal waters in April 2024, with dense traffic near major ports, also containing 10,000 vessel sequences and 2,000,000 records from 4,723 vessels, each lasting on average 2.8 hours. Both datasets include various vessel types (e.g., Cargo, Tanker, Passenger), providing comprehensive coverage for evaluation.

\subsubsection{Evaluation Metrics}
For longitude and latitude, we report axis-wise Mean Absolute Error (MAE), a robust measure of average error, and Root Mean Squared Error (RMSE), which emphasizes large deviations, both in degrees.
For the joint spatial error, we use the Haversine distance~\cite{wiki_haversine_2025} with Earth’s mean radius $R{=}6371$ km and report the Mean Haversine Distance (MHD) in kilometers.

\subsubsection{Baseline Methods} 
We compare \textsf{VISTA} with rule-based, deep learning-based, and LLM-based baselines. For rule-based methods, we include:
Lin-ITP~\cite{widyantara_improvement_2023} linearly interpolates between anchors, {Akima Spline}~\cite{zaman_interpolation-based_2023} yields visually smooth, $C^1$-continuous paths that capture gentle curvature, and {Kalman Filter}~\cite{wang_kinematic_2023} assumes a linear Gaussian state space with constant velocity and suits near-linear motion.
For deep learning-based methods, we include: {MH-GIN}~\cite{liu2025mhgin} models multi-scale features and cross-attribute dependencies on a heterogeneous graph to impute all attributes in AIS data, and {Multi-task AIS}~\cite{nguyen2018multi} is a recurrent framework with latent variables and AIS embeddings for noisy, irregular sampling.
For LLM-based methods, we include: {KAMEL}~\cite{musleh2023kamel} treats completion as masked infilling with spatially aware tokenization; we also assess out-of-the-box general models by using two families, Qwen~\cite{team_Qwen3_2025} and GLM 4.5~\cite{zeng2025glm}, each with a lightweight and a full-capacity variant with thinking mode enabled~\cite{Alibaba_Deep_2025}, yielding two pairs: {Qwen-plus-th} vs {Qwen-flash-th} and {GLM-4.5-air-th} vs {GLM-4.5-th}.

\subsubsection{Experimental Settings}
All experiments are conducted on a server with Intel Xeon Processor (Icelake) CPUs, 100\,GB RAM, and two NVIDIA A10 GPUs (each with 23\,GB memory). For data processing, each dataset is partitioned into 80\% training and 20\% testing. For comparability, all results reported below are computed on the held-out test split. To emulate realistic block missingness, trajectories are divided into minimal segments of length $m=20$, and each minimal segment is removed with probability $0.2$ (see Section~\ref{sec:problemdefinition}). The mask configuration is shared across all methods. The more detailed experimental setting is provided in Appendix~\ref{app:Detailed Experimental Setting}.
%--------------------------------------------------------

\subsection{Performance Evaluation}

\subsubsection{Effectiveness Analysis.}
As shown in Table~\ref{tab:main-1}, which reports the results on AIS-DK and AIS-US across MAE, RMSE, and MHD, boldface denotes the best performance, while underlining indicates the second-best. Building on these results, we summarize four observations.
1) \textsf{VISTA} consistently outperforms all baselines, improving over the strongest baseline, MH-GIN, by 5\%--91\%. These results demonstrate that grounding LLM reasoning in data-verified SDK yields substantially more accurate imputation than any single-paradigm approach.
2) All models perform better on AIS-DK, and the inter-model spread is smaller than on AIS-US, consistent with shorter trajectory durations and smaller inter-sample time intervals in AIS-DK, which produce shorter missing gaps and an easier imputation setting. 
3) Within the LLM group, the trajectory-oriented KAMEL performs markedly better than out-of-the-box general thinking-mode LLMs (Qwen-plus-th, Qwen-flash-th, GLM-4.5-th, GLM-4.5-air-th), underscoring the value of task-specific spatio-temporal tokenization and a ``physical grammar'' for trajectory imputation. Nevertheless, KAMEL still trails \textsf{VISTA} by a wide margin, highlighting the advantage of grounding LLM reasoning in a data-verified knowledge base (SD-KG). The general models also show that ``thinking'' alone is insufficient, since capability strongly affects accuracy: higher-capacity variants outperform their lighter counterparts (Qwen-plus-th and GLM-4.5-th outperform Qwen-flash-th and GLM-4.5-air-th), and in our setting, the GLM family tends to outperform its Qwen counterpart within each capacity tier (GLM-4.5-th vs.\ Qwen-plus-th, GLM-4.5-air-th vs.\ Qwen-flash-th), although all remain well below \textsf{VISTA}.
4) Among all baselines, MH-GIN achieves the best overall performance, followed closely by KAMEL and Multi-task AIS, reflecting the advantage of task-specific or data-driven modeling over rule-based and general-purpose LLM approaches. Nevertheless, MH-GIN still trails \textsf{VISTA} by a wide margin across all metrics.
Beyond these quantitative results, a case study in Appendix~\ref{app:casestudy} qualitatively illustrates how \textsf{VISTA} leverages SD-KG-driven reasoning to produce interpretable trajectory reconstructions with repair provenance.

% Building on this ranking, we summarize the key findings. 
% 1) \textsf{VISTA} consistently outperforms all baselines, improving over the strongest baseline by \warning{52.80\%--96.30\%}, evidencing effective mining of AIS domain knowledge and the productive use of underlying knowledge (structured domain priors plus inherent knowledge in LLMs) for trajectory imputation; 
% 2) within the LLM group, Qwen-flash-th and GLM-4.5-air-th are weaker than Qwen-plus-th and GLM-4.5-th, indicating that capacity matters inside the LLM family, yet even the stronger LLMs remain below classical rule-based methods under our out-of-the-box setting. This suggests that, although LLMs possess vast internet-scale knowledge, they lack sufficient maritime domain grounding and are not adept at precise numerical computation, which limits their performance on trajectory imputation.
% 3) deep learning-based methods (Multi-task AIS, MH-GIN, KAMEL) surpass rule-based methods (Lin-ITP, Cubic Spline, Kalman Filter), reflecting the advantage of data-driven modeling when sufficient training data are available; 
% 4) the gap ``LLMs < rule-based < learned < \textsf{VISTA}'' suggests that internet-scale pretraining alone is insufficient for trajectory imputation task, and that explicit domain grounding and structured priors -- central to \textsf{VISTA} -- are necessary to unlock strong performance of LLMs.

\subsubsection{Efficiency Analysis.}
As shown in Table~\ref{tab:timeOnly}, \textsf{VISTA} attains the lowest time cost, reducing time by 51\%--93\% relative to the thinking-mode LLM baselines. The gains stem from two complementary factors: first, the SD-KG compresses raw AIS into structured priors that filter noise and constrain candidate behavior patterns and imputation functions, enabling targeted retrieval and lightweight execution instead of open-ended LLM reasoning; second, the Workflow Manager Layer coordinates parallel knowledge extraction and imputation with scheduling, fault tolerance, and de-redundancy, stabilizing stochastic LLM latency and eliminating redundant information in the SD-KG.

% To further quantify deployment cost, Table~\ref{tab:cost-analysis} reports the per-segment LLM cost of \textsf{VISTA} alongside LLM baselines. Although \textsf{VISTA} issues multiple LLM calls per segment due to its multi-role workflow, the SD-KG constrains each call to a targeted selection from pre-filtered candidate sets, rather than open-ended, end-to-end reasoning from boundary coordinates as in the baselines. This yields shorter per-call reasoning chains and lower overall token consumption and cost. Overall, \textsf{VISTA} achieves a \warning{XX\%--XX\%} lower per-segment cost compared to the thinking-mode baselines, confirming that the SD-KG-driven decomposition not only improves accuracy but also reduces economic overhead.
% \input{src/Tables/5-CostAnalysis}

We also evaluate the reliability of the Imputation Method Builder's code generation. As shown in Table~\ref{tab:code-reliability}, the first-pass success rate is 64.3\% on AIS-DK and 29.5\% on AIS-US. With the built-in verification-retry mechanism ($\varrho_r{=}3$), the success rate climbs steadily over successive attempts, reaching 77.4\% and 70.2\% after three attempts with an average of only 0.61 and 1.02 retries per segment, respectively. The lower first-pass rate on AIS-US reflects its longer trajectories and more complex traffic patterns near major U.S.\ coastal ports, which demand more sophisticated imputation functions; nevertheless, the retry mechanism narrows this gap substantially.
\begin{table}[!tbp]
  \centering
  \small
  \caption{Cumulative success rate (\%) of LLM-generated imputation code over successive verification-retry attempts.}
  \label{tab:code-reliability}
  \setlength{\tabcolsep}{10pt}
  \begin{tabular}{lccc c}
    \toprule
    \multirow{2}{*}{Dataset} & \multicolumn{3}{c}{Attempt} & \multirow{2}{*}{\makecell{Avg.\\Retries}} \\
    \cmidrule(lr){2-4}
    & 1\textsuperscript{st} & 2\textsuperscript{nd} & 3\textsuperscript{rd} & \\
    \midrule
    \textbf{AIS-DK} & 64.3 & 73.7 & 77.4 & 0.61 \\
    \textbf{AIS-US} & 29.5 & 67.1 & 70.2 & 1.02 \\
    \bottomrule
  \end{tabular}
  % \vspace{-4mm}
\end{table}

\subsubsection{Scalability Analysis.}
To evaluate how \textsf{VISTA} scales with data size, we run end-to-end experiments on subsets of 2{,}000, 5{,}000, and 10{,}000 trajectories for both AIS-DK and AIS-US.
As shown in Table~\ref{tab:scalability}, the two datasets exhibit distinct scaling behaviors.
On AIS-US, all five metrics remain nearly constant across the three scales (MHD varies within 1.41--1.45km), indicating that the SD-KG captures sufficient behavioral knowledge even from 2{,}000 trajectories; adding more data introduces neither improvement nor degradation, since the dense and repetitive port traffic near major U.S.\ coastal hubs provides adequate pattern coverage at smaller scales.
On AIS-DK, performance is non-monotonic: MHD slightly increases from 0.34 at 2{,}000 to 0.40 at 5{,}000 before dropping to 0.24 at 10{,}000.
This pattern suggests that at intermediate scales, the SD-KG absorbs more diverse but potentially conflicting behavior patterns from the heterogeneous Baltic and North Sea routes, temporarily increasing noise in candidate retrieval; as data grows further to 10{,}000, the statistical support for each pattern strengthens, enabling the knowledge base to disambiguate among candidates and produce more accurate imputation.
Overall, \textsf{VISTA} maintains robust performance across all tested scales, confirming that the SD-KG construction and retrieval pipeline generalizes well without dataset-specific tuning.
\begin{table}[!tbp]
  \centering
  \small
  \caption{Scalability analysis with varying data sizes.}
  % \vspace{-2mm}
  \label{tab:scalability}
  \setlength{\tabcolsep}{2.5pt}
  \begin{tabular}{l|c|ccccc}
    \toprule
    Dataset & \makecell{Trajec-\\tories} & MAE(Lat) & RMSE(Lat) & MAE(Lon) & RMSE(Lon) & MHD \\
    \midrule
    \multirow{3}{*}{\textbf{AIS-DK}}
      & 2{,}000  & 2.332e-3 & 5.838e-2 & 2.389e-3 & 4.988e-2 & 0.3435 \\
      & 5{,}000  & 2.850e-3 & 6.945e-2 & 2.247e-3 & 5.430e-2 & 0.3950 \\
      & 10{,}000 & 1.817e-3 & 4.324e-2 & 1.123e-3 & 4.027e-2 & 0.2418 \\
    \midrule
    \multirow{3}{*}{\textbf{AIS-US}}
      & 2{,}000  & 7.504e-3 & 2.194e-2 & 1.035e-2 & 3.797e-2 & 1.4079 \\
      & 5{,}000  & 7.508e-3 & 1.968e-2 & 1.013e-2 & 2.972e-2 & 1.4310 \\
      & 10{,}000 & 7.587e-3 & 2.050e-2 & 1.027e-2 & 2.942e-2 & 1.4532 \\
    \bottomrule
  \end{tabular}
  % \vspace{-4mm}
\end{table}

% We further conduct a robustness evaluation under different missing ratios to assess performance stability, as detailed in Appendix~\ref{app:addtionalExp}.

\subsection{Ablation Study}
We conduct two ablation studies.
i) LLM role ablation: We vary the LLM capacity across three roles, Analysis (Behavior Abstraction and Explanation Composer), Programming (Method Builder), and Decision (Behavior Estimator and Method Selector), by downgrading one role while keeping the others fixed. Table~\ref{tab:ablation-combined} reports the results on AIS-DK and AIS-US.
ii) Workflow ablation: We assess the workflow management layer under different batch sizes and with or without the De-redundancy (DR) mechanism. Table~\ref{tab:ablation-managelayer} presents the execution time and the total number of behavior pattern and imputation method nodes in the SD-KG (denoted as Size).

% From Table~\ref{tab:ablation-combined}, Analysis and Programming are capacity sensitive, and lighter models noticeably degrade MAE, RMSE, and MHD due to heavier semantic abstraction and method synthesis. Decision is more tolerant, benefiting from strong priors supplied by the SD-KG, which simplifies candidate scoring and selection. 
% From Table~\ref{tab:ablation-managelayer}, per-instance latency remains stable as batch size increases, indicating linear throughput scaling without contention. Disabling De-redundancy increases time cost by 1.89\% (AIS-DK, batch 5), 2.19\% (AIS-DK, batch 9), 1.70\% (AIS-US, batch 5), and 2.15\% (AIS-US, batch 9), confirming that De-redundancy is effective under multi-batch settings. Overall, \textsf{VISTA} attains accuracy through SD-KG guided reasoning and preserves scalability via workflow level parallelism and redundancy control.
From Table~\ref{tab:ablation-combined}, the Decision role is the most capacity-sensitive: downgrading it from Qwen-plus to Qwen-flash increases MHD by roughly 12$\times$ on both datasets, showing that accurate behavior estimation and method selection demand stronger reasoning and contextual inference. Programming degrades moderately (MHD increases by about 2$\times$), as generating valid imputation code still benefits from the structured priors in the SD-KG but requires sufficient coding capability. Analysis is the most tolerant: MHD increases by only about 10\%, indicating that behavior abstraction and explanation composition can largely rely on the SD-KG's structured vocabulary and templates even with a lighter model.
From Table~\ref{tab:ablation-managelayer}, throughput scales almost linearly with batch size, confirming that the workflow layer achieves efficient and contention-free parallel scheduling. Disabling de-redundancy slightly shortens runtime but significantly increases SD-KG size (e.g., $1,105 \rightarrow 1,400$ on AIS-DK and $2,057 \rightarrow 4,109$ on AIS-US), indicating a trade-off between speed and compact SD-KG. 
% In other words, DR introduces limited overhead to maintain richer SD-KG content, ensuring that each workflow iteration produces more comprehensive behavioral patterns and imputation methods. This balance allows \textsf{VISTA} to preserve both computational efficiency and interpretability at scale.

\subsection{Robustness Analysis}
We evaluate \textsf{VISTA}'s robustness under two dimensions of data degradation: longer missing gaps (varying segment length $m$) and higher data loss (varying missing ratio). Both dimensions are external data conditions rather than internal design choices, and together they stress-test how well the SD-KG-grounded pipeline tolerates increasingly difficult imputation scenarios.

\subsubsection{Varying Segment Length.}
Table~\ref{tab:segment-length} reports performance under $m \in \{10, 20, 40\}$ with a fixed missing ratio of 20\%.
On AIS-DK, MHD increases from 0.06 at $m{=}10$ to 0.24 at $m{=}20$ and 0.59 at $m{=}40$, remaining below 1km even when each missing gap spans 40 consecutive time steps. Notably, at $m{=}10$ the MAE values drop to the $10^{-4}$-degree range (sub-100m accuracy), confirming that the SD-KG can almost perfectly reconstruct short gaps where boundary anchors tightly constrain the imputation.
On AIS-US, performance is stable from $m{=}10$ to $m{=}20$ (MHD~1.40--1.45km) and increases moderately at $m{=}40$ (MHD~2.46km, a 69\% rise from $m{=}20$), still remaining below 3km.
The wider gap between AIS-DK and AIS-US at $m{=}40$ is consistent with the dataset characteristics reported in Section~\ref{sec:experiment}: AIS-DK trajectories average 0.5~hours with relatively short Baltic and North Sea routes, so even 40-step gaps remain within the SD-KG's pattern coverage; AIS-US trajectories average 2.8~hours with diverse maneuvers near major ports, giving the LLM a wider spatial extent to extrapolate over longer gaps.
\begin{table}[!tbp]
  \centering
  \small
  \caption{Imputation performance under varying segment lengths $m \in \{10, 20, 40\}$ with a fixed missing ratio of 20\%.}
  % \vspace{-2mm}
  \label{tab:segment-length}
  \setlength{\tabcolsep}{2.5pt}
  \begin{tabular}{l|c|ccccc}
    \toprule
    Dataset & $m$ & MAE(Lat) & RMSE(Lat) & MAE(Lon) & RMSE(Lon) & MHD \\
    \midrule
    \multirow{3}{*}{\textbf{AIS-DK}}
      & 10 & 3.720e-4 & 5.264e-4 & 4.960e-4 & 7.980e-4 & 0.0571 \\
      & 20 & 1.817e-3 & 4.324e-2 & 1.123e-3 & 4.027e-2 & 0.2418 \\
      & 40 & 4.536e-3 & 8.908e-2 & 2.465e-3 & 4.683e-2 & 0.5949 \\
    \midrule
    \multirow{3}{*}{\textbf{AIS-US}}
      & 10 & 7.188e-3 & 1.899e-2 & 9.969e-3 & 2.965e-2 & 1.3959 \\
      & 20 & 7.587e-3 & 2.050e-2 & 1.027e-2 & 2.942e-2 & 1.4532 \\
      & 40 & 1.221e-2 & 3.910e-2 & 1.887e-2 & 6.826e-2 & 2.4607 \\
    \bottomrule
  \end{tabular}
\end{table}

\subsubsection{Varying Missing Ratio.}
Table~\ref{tab:missingratio-1} reports performance under missing ratios of $\{10\%, 20\%, 30\%\}$ with a fixed segment length $m{=}20$.
On AIS-DK, \textsf{VISTA} is highly robust: MHD increases by only 17\% from 0.22 at 10\% to 0.25 at 30\%, and MAE values remain nearly flat across all three ratios (e.g., MAE (Lat) varies within $1.3\text{e-}3$--$2.1\text{e-}3$), indicating that per-segment imputation quality is largely unaffected by the proportion of missing data.
On AIS-US, MHD rises steadily from 1.19 at 10\% to 1.45 at 20\% and 1.70 at 30\%, a 42\% total increase that remains moderate in absolute terms ($<2$km).
Across both robustness experiments, \textsf{VISTA} maintains stable performance: AIS-DK remains within sub-kilometer accuracy under all tested conditions, while AIS-US shows gradual degradation that tracks the increasing difficulty but never exhibits catastrophic failure. The larger sensitivity on AIS-US reflects its longer trajectories and more complex traffic patterns, suggesting that enriching the SD-KG with more diverse behavioral patterns---e.g., through larger training sets or cross-region knowledge transfer---could further tighten the gap.
\begin{table}[!tbp]
  \centering
  \small
  \caption{Imputation performance under varying missing ratios $\{10\%, 20\%, 30\%\}$ with a fixed segment length $m{=}20$.}
  \label{tab:missingratio-1}
  \setlength{\tabcolsep}{1.5pt}
  \begin{tabular}{l|c|ccccc}
    \toprule
    Dataset & \makecell{Missing\\Ratio} & MAE(Lat) & RMSE(Lat) & MAE(Lon) & RMSE(Lon) & MHD \\
    \midrule
    \multirow{3}{*}{\textbf{AIS-DK}}
      & 10\% & 1.338e-3 & 3.653e-2 & 1.092e-3 & 3.114e-2 & 0.2167 \\
      & 20\% & 1.817e-3 & 4.324e-2 & 1.123e-3 & 4.027e-2 & 0.2418 \\
      & 30\% & 2.136e-3 & 7.138e-2 & 2.555e-3 & 1.298e-1 & 0.2529 \\
    \midrule
    \multirow{3}{*}{\textbf{AIS-US}}
      & 10\% & 6.626e-3 & 2.001e-2 & 8.630e-3 & 2.953e-2 & 1.1932 \\
      & 20\% & 7.587e-3 & 2.050e-2 & 1.027e-2 & 2.942e-2 & 1.4532 \\
      & 30\% & 8.647e-3 & 2.555e-2 & 1.243e-2 & 3.913e-2 & 1.6977 \\
    \bottomrule
  \end{tabular}
  % \vspace{-5mm}
\end{table}

To further illustrate how \textsf{VISTA} produces interpretable repair provenance for individual trajectory segments, we provide a detailed case study in Appendix~\ref{app:casestudy}. We have also developed \texttt{CLEAR}~\cite{liu2025clear}, an interactive demo platform built on \textsf{VISTA} that enables non-expert users to explore, complete, and annotate vessel trajectories through a knowledge-graph-driven interface.

\section{Conclusion}
\label{sec:conclusion}
We present \textsf{VISTA}, a knowledge-driven vessel trajectory imputation framework that equips repaired trajectories with structured repair provenance---queryable metadata recording what behavior pattern is estimated, how the segment is imputed, and why the behavior occurred.
By grounding LLM reasoning in a data-verified Structured Data-derived Knowledge Graph (SD-KG) through a closed data--knowledge--data loop, \textsf{VISTA} ensures that behavior estimation, imputation method selection, and causal explanation generation are all anchored in empirically validated knowledge rather than unconstrained LLM conjecture.
A workflow management layer coordinates the full pipeline with parallel scheduling, fault tolerance, and redundancy control, enabling efficient processing of large-scale AIS data.
Experiments on two real-world AIS datasets show that \textsf{VISTA} outperforms the strongest baseline by 5\%--91\% in accuracy while reducing inference time by 51\%--93\%; ablation studies confirm that each LLM role contributes distinctively and that de-redundancy maintains a compact yet expressive SD-KG.
In future work, we plan to extend \textsf{VISTA} into a unified maritime trajectory management framework supporting trajectory imputation, anomaly detection, and trajectory prediction.

\section{Acknowledgments}
This research was supported in part by the European Union-funded Project MobiSpaces under grant agreement no 101070279.
% \section{Acknowledgment}
% The research leading to the results presented in this paper has received funding from the European Union-funded Project MobiSpaces under grant agreement no 101070279.
% ais.app.cs.aau.dk/map

\balance
\bibliographystyle{ACM-Reference-Format}
\bibliography{reference}

\clearpage
\appendix 
\section{The details of \textsf{VISTA}}
\subsection{The SD-KG Construction and Maintenance} \label{app:M1-M3}
\begin{figure*}[!ht]
\centering
\begin{tcolorbox}[
  colback=gray!8, colframe=black!80,
  boxrule=0.5pt, arc=3mm,
  left=8pt, right=8pt, top=6pt, bottom=6pt,   
  title=Behavior Abstraction,
  colbacktitle=gray!40, coltitle=black, fonttitle=\bfseries
]
\small
[\textbf{TASK}]\\
You are \textbf{an expert in maritime data analysis}.\\
Your task is to generate a list of specific, interpretable patterns that describe how both \textbf{latitude and longitude} (vessel positions) can be inferred from a set of AIS features. \\
These patterns will be used to \textbf{impute missing values of latitude and longitude} in AIS data. Each pattern must be: 
\begin{itemize}[leftmargin=15pt,itemsep=1pt,topsep=1pt,parsep=0pt]
    \item \textbf{Concrete and usable}: describing a clear condition on the target features and the corresponding position range;
    \item \textbf{Simultaneous}: including both latitude and longitude ranges, or describing a \textbf{trajectory pattern} (e.g., a curved path, straight line, or repeated loop);
    \item \textbf{Explainable}, with a short justification after each pattern explaining why this condition relates to the given position;
    \item \textbf{Mathematically expressive}: including a possible trajectory equation or shape that approximates the vessel’s movement under this condition.
\end{itemize}
[\textbf{INPUT}]\\
You are given by: 
\begin{itemize}[leftmargin=15pt,itemsep=1pt,topsep=1pt,parsep=0pt]
  \item A sample of trajectory data (latitude, longitude, and various AIS features): \texttt{\{trajectory\_data\}}
\end{itemize}
[\textbf{OUTPUT}]\\
Please strictly follow the following format, return only one pattern,and output all patterns in triple quotes: \\
\verb|'''|\\
Pattern: 
\begin{itemize}[leftmargin=15pt,itemsep=1pt,topsep=1pt,parsep=0pt]
\item \texttt{speed\_pattern}: speed profile without numerical values and punctuation (detailed description for speed profile).
\item \texttt{course\_pattern}: change in course over ground without numerical values and punctuation (detailed description for change in course over ground).
\item \texttt{heading\_pattern}: heading fluctuation without numerical values and punctuation (detailed description for heading fluctuation).
\item \texttt{intent}: inferred maneuver intention without numerical values and punctuation (detailed description for inferred maneuver intention).
\end{itemize}
\verb|'''|\\
For \texttt{speed\_pattern}, You can choose from \texttt{\{speed\_dict\}}, and if you don't have a suitable one, you can create a new one. \\
For \texttt{course\_pattern}, You can choose from \texttt{\{course\_dict\}}, and if you don't have a suitable one, you can create a new one. \\
For \texttt{heading\_pattern}, You can choose from \texttt{\{heading\_dict\}}, and if you don't have a suitable one, you can create a new one. \\
For \texttt{intent}, You can choose from \texttt{\{intent\_dict\}}, and if you don't have a suitable one, you can create a new one. \\
\vspace{0em}
[\textbf{EXAMPLE}]\\
% \small\ttfamily
\verb|'''|\\
Pattern: \\
- \texttt{speed\_pattern}: stable (the vessel is maintaining a consistent speed, not accelerating or decelerating)\\
- \texttt{course\_pattern}: stable (the vessel is maintaining a consistent course over ground)\\
- \texttt{heading\_pattern}: stable (the heading does not fluctuate significantly, indicating no sharp maneuvers)\\
- \texttt{intent}: navigating (the vessel is maintaining its course)\\
\verb|'''|

Make sure that your output strictly follows this format. Any patterns that do not adhere to this structure should be adjusted to fit the template. If any pattern involves multiple segments, please aggregate them.
\end{tcolorbox}
% \vspace{-1mm}
\caption{Prompt of behavior abstraction.}
% \vspace{-1mm}
\label{fig:prompt-behaviorabstraction}
\end{figure*}

\subsubsection{Behavior Pattern Extraction} 
\label{prompt-BehaviorPatternsExtraction}

Figure~\ref{fig:prompt-behaviorabstraction} presents the Behavior Abstraction Prompt, which operationalizes the behavior abstraction in Section~\ref{sec:ku-extraction} by mapping raw motion signals into a discrete, human-interpretable pattern for downstream imputation. 
The kinematic tokens $(p^s,p^\theta,p^\psi)$ are instantiated as \texttt{speed\_pattern}, \texttt{course\_pattern}, and \texttt{heading\_pattern}, respectively. The prompt explicitly instructs the LLM to analyze the temporal evolution within the provided \{\texttt{trajectory\_data}\} and to choose from finite vocabularies (\{\texttt{speed\_dict}\}, \{\texttt{course\_dict}\}, \{\texttt{heading\_dict}\}) before creating any new entries, thereby constraining vocabulary growth and enabling deduplication of semantically equivalent labels.
The intent token $p^i$ is realized through two complementary fields: a high-level intent (e.g., `navigating') and a context-grounded navigation intent (e.g., `the vessel is maintaining its course'). 
Both are inferred from static and contextual cues, navigation status $\eta$, vessel type $\kappa$, cargo type $\chi$, and spatial context $\sigma$ (obtained from Static and Spatial Encoder) in \texttt{trajectory\_data}; a finite set \{\texttt{intent\_dict}\} regulates label usage to preserve interpretability and retrieval efficiency.
Finally, the output schema mandates a single structured pattern enclosed in triple quotes, ensuring robust parsing and consistent downstream indexing.

\begin{figure*}[t]
\centering
\begin{tcolorbox}[
  colback=gray!8, colframe=black!80,
  boxrule=0.5pt, arc=3mm,
  left=8pt, right=8pt, top=6pt, bottom=6pt,   % 给行号留出左边距
  title=Imputation Method Builder,
  colbacktitle=gray!40, coltitle=black, fonttitle=\bfseries
]
\small
[\textbf{TASK}]
You are an expert in maritime trajectory analysis and spatio-temporal modeling, specialized in developing interpretable algorithms for vessel movement reconstruction. You are given vessel trajectory data and are asked to generate a \texttt{spatial\_function} that estimates missing latitude and longitude positions based on known trajectory features and motion patterns. 

You need to adhere requirements: 
\begin{itemize}[leftmargin=10pt,itemsep=1pt,topsep=1pt,parsep=0pt]
\item The \texttt{spatial\_function} should be a single-line, directly executable Python function. 
\item The \texttt{spatial\_function} is encouraged to choose from a variety of path models, not just linear interpolation. 
\item The \texttt{spatial\_function} must compute a sequence of intermediate points using the provided \texttt{start}, \texttt{end} and \texttt{Time\_interval}.
\begin{itemize}[leftmargin=10pt,itemsep=1pt,topsep=1pt,parsep=0pt]
  \item \texttt{start}: A tuple representing the geographic coordinates (latitude, longitude) immediately before the missing data block.\\
  Type: \texttt{Tuple[float, float]}
  \item \texttt{end}: A tuple representing the geographic coordinates (latitude, longitude) immediately after the missing data block. \\
  Type: \texttt{Tuple[float, float]}
  \item \texttt{Time\_interval}: A list of time differences in seconds relative to the timestamp of the starting point, covering the entire range including the point before and after the missing block. \\
  Type: \texttt{List[float]} 
\end{itemize}
\end{itemize}

[\textbf{INPUT}]
You are given by: 
\begin{itemize}[leftmargin=10pt,itemsep=1pt,topsep=1pt,parsep=0pt]
\item trajectory: \texttt{\{trajectory\_data\}}.
\item behavior pattern of trajectory: \texttt{\{pattern\}}.
\end{itemize}

[\textbf{OUTPUT}]
Please strictly follow the following format: \\
\textbf{Function:}
\small\ttfamily
\verb|'''|
def spatial\_function(start, end, Time\_interval): return [...] 
\verb|'''|
\normalfont

\textbf{Description:} A brief explanation of what this function does, including how it uses the input parameters.

[\textbf{EXAMPLE}]\\
\small\ttfamily
\verb|'''|def spatial\_function(start, end, Time\_interval): return []\verb|'''|
\normalfont

\texttt{\{feedback\_text\_description\}}
\end{tcolorbox}
% \vspace{-1mm}
\caption{Prompt of imputation method builder.}
% \vspace{-1mm}
\label{fig:function-prompt-maritime}
\end{figure*}

\begin{figure*}[t]
\centering
\begin{tcolorbox}[
  colback=gray!8, colframe=black!80,
  boxrule=0.5pt, arc=3mm,
  left=8pt, right=8pt, top=6pt, bottom=6pt,   % 给行号留出左边距
  title=Behavior Selection and Rationale,
  colbacktitle=gray!40, coltitle=black, fonttitle=\bfseries
]
\small
[\textbf{TASK}] 
You are an \textbf{expert maritime behavior analyst}, specialized in interpreting vessel movement patterns and reasoning over graph-based representations of AIS knowledge. You need to select the most plausible \textbf{movement (behavior pattern)} for the current gap. Specifically, the process is as follow: 
\begin{itemize} [leftmargin=14pt,itemsep=1pt,topsep=1pt,parsep=0pt]
  \item Analyze the boundary movement patterns and DOT graph structure, then rely on their evidence weights to shortlist \textbf{Top-\{top\_k\}} movements.
  \item Choose \textbf{ONE} final movement ID for the gap.
  \item Provide a two-part rationale:
    \begin{enumerate}[leftmargin=14pt,itemsep=0.5pt,topsep=0.5pt,parsep=0pt]
      \item \textbf{Graph Support}: cite the most informative vessel$\rightarrow$movement edges (IDs/weights) that support your choice.
      \item \textbf{Contextual Justification}: explain consistency with boundary movement patterns ($v^{-}_b$, $v^{+}_b$) and the gap's boundary conditions.
    \end{enumerate}
  \item Output in the following block (\textbf{no extra text}):
\end{itemize}

[\textbf{INPUT}]
You are given by: 
\begin{itemize}[leftmargin=12pt,itemsep=1pt,topsep=1pt,parsep=0pt]
  \item Boundary movement patterns (behavior patterns extracted from adjacent segments on both sides of the missing block):
  {\footnotesize\ttfamily \{boundary\_text\}}
  \item Induced subgraph in DOT (vessel$\rightarrow$movement and movement$\rightarrow$function edges with weights):
  {\footnotesize\ttfamily \{dot\_text\}}
  \item Candidate movements (with tokens, graph priors and used edges):
  {\footnotesize\ttfamily \{movement\_text\}}
  \item Contextual static attributes inferred from neighboring segments (vessel nodes):
  {\footnotesize\ttfamily \{context\_vessels\}}
\end{itemize}

\medskip
[\textbf{OUTPUT}]
Please strictly follow the following format:\\
\medskip
{\small\ttfamily
\verb|'''|\\
Selected Movement ID: <ID> \\
Graph Support: <edges and weights you rely on> \\
Contextual Justification: <why consistent with boundary context> \\
\verb|'''|
}

\end{tcolorbox}
% \vspace{-1mm}
\caption{The prompt of behavior selection and rationale.}
% \vspace{-1mm}
\label{fig:behavior-estimator-prompt}
\end{figure*}
%A.1.2
\subsubsection{Imputation Method Builder} \label{prompt-ImputationFunctionGeneration}

Figure~\ref{fig:function-prompt-maritime} illustrates the prompt design of the Imputation Method Builder for imputation method generation in \textsf{VISTA}. It instantiates the imputation method $v_f=(f,d(f))$ defined in Section~\ref{sec:methodbuilder} by compelling the LLM to output a single-line, directly executable Python function $f$ together with a concise textual description $d(f)$. The prompt enforces a fixed, machine-readable function format and explicitly defines the meaning of each input variable, where \texttt{start} and \texttt{end} denote the boundary coordinates of the missing segment and \texttt{Time\_interval} specifies the temporal span across both sides of the gap. Contextual grounding is provided through \texttt{trajectory\_data} and the behavior pattern \texttt{\{pattern\}}, guiding the generator to incorporate pattern-specific motion characteristics during function synthesis. By explicitly allowing non-linear paths (``not just linear interpolation''), the prompt expands the hypothesis space while maintaining direct executability and reproducibility via a unified output schema (triple-quoted code block plus a separate description). 

\begin{algorithm}[!h]
\caption{SD-KG Construction and Incremental Update}
\small
\label{alg:sdk}
\KwInput{AIS dataset $\mathcal{X}$; thresholds $m$, $\varrho_f$ and $\varrho_r$.}
\KwOutput{$\mathcal{G}_d=(\mathcal{V}_s,\mathcal{V}_b,\mathcal{V}_f,\mathcal{E}_{sb},\mathcal{E}_{bf})$, $\mathcal{U}_{d}$}
Initialize empty $\mathcal{G}_{d}$ and $\mathcal{U}_{d}$ \\
\ForEach{each vessel $\iota$ in $\mathcal{X}$}{
    $\langle \mathbf{S}_\iota^1,\dots,\mathbf{S}_\iota^K\rangle \;\gets\; \textsc{Partition}(\mathbf{X}_\iota; m)$\\
    $\mathbf{M}_\iota \;\gets\; \textsc{GetSegmentMask}(\langle \mathbf{S}_\iota^1,\dots,\mathbf{S}_\iota^K\rangle)$\\
    $\mathcal{U}_{d,\iota} \leftarrow \langle \rangle$\\
    \ForEach{$k=1$ to $K$}{
        \uIf{$M_\iota^k=0$}{
            $\mathcal{U}_{d,\iota} \leftarrow \mathcal{U}_{d,\iota} \| \varnothing $\\
            \textbf{Continue}\\
        }
        % \bluecomment{Knowledge Unit Extraction (§\ref{sec:ku-extraction})}\\
        \bluecomment{Extract Static Attributes}\\
        $\iota,\eta,\chi,\kappa \;\gets\; \textsc{Mode}(\{x.\iota, x.\eta, x.\chi, x.\kappa \;|\; x\in \mathbf{S}_\iota^k\})$\\
        $\tilde d,\tilde \ell,\tilde \beta \;\gets\; \textsc{Discrete}(\{x.d,\ x.\ell,\ x.\beta \;|\; x\in \mathbf{S}_\iota^k\ \})$\\
        $\sigma \gets \textsc{Mode}(\{\textsc{Overpass}(x.\lambda, x.\phi) \;|\; x\in \mathbf{S}_\iota^k\})$\\
        $v_s \gets \{ \iota,\eta,\chi,\tilde d,\tilde \ell,\tilde \beta,\sigma,\kappa \}$\\
        \bluecomment{Extract Behavior Pattern)}\\
        % $p^s,p^\theta,p^\psi \gets \textsc{FeatExtwithDedup}(\{x.s, x.\theta, x.\psi \;|\; x\in \mathbf{S}_\iota^k\}, \mathcal{P}^s,\mathcal{P}^\theta,\mathcal{P}^\psi)$\\
        % $p^i \gets \textsc{IntentInferwithDedup} ((v_s,\ p^s,p^\theta,p^\psi,p^\tau);\ \mathcal{P}^i)$\\
        $p^s,p^\theta,p^\psi, p^i, p^{\tau}\gets \textsc{BehaviorAbstraction}(\{x \;|\; x\in \mathbf{S}_\iota^k\}, \mathcal{P}^s,\mathcal{P}^\theta,\mathcal{P}^\psi, \mathcal{P}^i)$\\
        $v_b \gets (p^s, p^\theta, p^\psi, p^i, p^\tau)$\\
        \bluecomment{Generate Imputation Method}\\
        $v_f \gets \textsc{RetrieveBest}(v_s,v_b,\mathcal{V}_f)$\\
        \lIf{$v_f =\varnothing$}{
            % $v_f \gets (\textsc{GenFunc}(\varnothing, \varnothing, v_s,v_b))$
            $v_f \gets \textsc{GenFunc}(v_s,v_b)$
        }
        \ForEach{$r=1$ to $\varrho_r$}{
            $e \gets \textsc{EvalMAE}(v_f.f,\mathbf{S}_\iota^k)$\\
            \lIf{$e\le\varrho_f$}{
                \textbf{break} 
            }
            \lElse {
                % $v_f \gets (\textsc{GenFunc}(f,e,v_s,v_b))$
                $v_f \gets \textsc{GenFunc}(v_s,v_b)$
            }
        }
        % \lIf{$v.d(f)=\varnothing$}{
        %     $v_f.d(f)\gets\textsc{Describe}(f,v_b)$
        % }
        \bluecomment{Update SD-KG}\\
        $\mathcal{V}_s, \mathcal{V}_b, \mathcal{V}_f \gets \mathcal{V}_s \cup v_s, \mathcal{V}_b \cup \{v_b\}, \mathcal{V}_f \cup \{v_f\}$\\
        $\mathcal{E}_{sb},\mathcal{E}_{bf} \gets \textsc{EdgeInc} ((v_s,v_b,v_f), \mathcal{E}_{sb},\mathcal{E}_{bf})$\\
        $\mathcal{U}_{d,\iota} \leftarrow \mathcal{U}_{d,\iota} \| (v_s,v_b,v_f)$\\
    }
    $\mathcal{U}_d \leftarrow \mathcal{U}_d \bigcup \{ \mathcal{U}_{d,\iota}\}$\\
}
\KwRet $\mathcal{G}_d=(\mathcal{V}_s,\mathcal{V}_b,\mathcal{V}_f,\mathcal{E}_{sb},\mathcal{E}_{bf})$ , $\mathcal{U}_d$\;
\end{algorithm}
\subsubsection{Overall Procedure of SD-KG Construction and Maintenance}
\label{sec:alg-sdk}
Algorithm~\ref{alg:sdk} summarizes the end-to-end pipeline that builds and updates the SD-KG from complete minimal segments. Line 1 initializes the empty graph \(\mathcal{G}_d\) and the global per-vessel unit collection \(\mathcal{U}_d\). 
Lines 2--5 segment each vessel's stream (\textsc{Partition}) and build the mask (\textsc{GetSegmentMask}); then initialize the per-vessel knowledge unit sequence \(\mathcal{U}_{d,\iota}\gets\langle\rangle\), which records, in segment order, the knowledge unit built for each segment. 
Lines 6--9 skip incomplete segments: if \(M_\iota^k=0\), append \(\varnothing\) to \(\mathcal{U}_{d,\iota}\) as a placeholder for a missing unit and continue; this preserves alignment between segment index \(k\) and entries in \(\mathcal{U}_{d,\iota}\). 
Lines 10--15 extract static attributes: \textsc{Mode} selects segment-wise majorities for categorical fields, \textsc{Discrete} bins draught and geometry (Section~\ref{sec:detailKnowledgeBase}), and \textsc{Overpass} yields the spatial context; these form \(v_s\). 
Lines 16--17 obtain the behavior pattern (Appendix~\ref{prompt-BehaviorPatternsExtraction}); together they form \(v_b\). 
Lines 18--24 pick an imputation method \(v_f\): \textsc{RetrieveBest} reuses a candidate from \(\mathcal{V}_f\) using context and edge statistics (Section~\ref{sec:InterpretableImputation}); otherwise \textsc{GenFunc} (Appendix~\ref{prompt-ImputationFunctionGeneration}) generates executable code \(f\). The loop validates with \textsc{EvalMAE} (mean of latitude/longitude MAE) against \(\varrho_f\), refining via \textsc{GenFunc} up to \(\varrho_r\) times. 
Lines 25--28 integrate results into SD-KG: nodes are inserted or reused, and \textsc{EdgeInc} creates or increments \((v_s,v_b)\) and \((v_b,v_f)\) edges to support later retrieval; append the knowledge unit \((v_s,v_b,v_f)\) to \(\mathcal{U}_{d,\iota}\) at position \(k\).
Line 29 aggregates each vessel's knowledge unit sequence into the global collection: \(\mathcal{U}_d \leftarrow \mathcal{U}_d \cup \{\mathcal{U}_{d,\iota}\}\). Thus, \(\mathcal{U}_d\) is a vessel-indexed collection of sequences where each \(\mathcal{U}_{d,\iota}\) preserves segment order and may contain \(\varnothing\) at masked positions.
Line 30 returns the updated SD-KG \(\mathcal{G}_d\) and the knowledge unit collection \(\mathcal{U}_d\).

% Algorithm~\ref{alg:sdk} summarizes the end-to-end pipeline that builds and updates the SD-KG from complete minimal segments. Lines 1--4 initialize an empty graph and segment each vessel’s stream, then build the mask. Lines 5--7 skip incomplete segments. Lines 8--13 extract static attributes: \textsc{Mode} selects the segment-wise majority for categorical fields, \textsc{Discrete} bins draught and geometry as defined in Section~\ref{sec:detailKnowledgeBase}, and \textsc{Overpass} yields the spatial context; these form $v_s$. Lines 14--15 obtain the behavior pattern (see Appendix~\ref{prompt-BehaviorPatternsExtraction}); together they form $v_b$. Lines 16--23 pick an imputation method: \textsc{RetrieveBest} reuses a candidate from $\mathcal{V}_f$ using context and edge statistics (details in Section~\ref{sec:InterpretableImputation}); otherwise \textsc{GenFunc} (see Appendix~\ref{prompt-ImputationFunctionGeneration}) generates executable code $f$ and the loop validates it with \textsc{EvalMAE} (mean of latitude and longitude MAE) against $\vartheta_f$, refining via \textsc{GenFunc} up to $\vartheta_r$ times; \textsc{Describe} attaches $d(f)$ with intended use, assumptions, and parameters. Lines 24--26 integrate results: nodes are inserted or reused, and \textsc{EdgeInc} creates or increments weights for $(v_s,v_b)$ and $(v_b,v_f)$, providing empirical support for later retrieval. Line 27 returns the updated SD-KG $\mathcal{G}_d$.

\subsection{Knowledge-Driven Trajectory Imputation}
\label{app:knowledgedriventrajectoryimputation}
%A.2.1

\subsubsection{Behavior Pattern Selection and Rationale} \label{prompt-BehaviorSelection}

Figure~\ref{fig:behavior-estimator-prompt} presents the prompt used for LLM-driven Behavior Pattern Selection and Rationale described in Section~\ref{sec:behaviorEstimator}. This prompt guides the LLM to identify the most plausible behavior pattern for a missing trajectory segment and to articulate the reasoning behind this choice.

To achieve this, the prompt provides four structured inputs: i) boundary behavior patterns $v_b^{-}$ and $v_b^{+}$ extracted from adjacent trajectory segments (\texttt{\{boundary\_text\}}), ii) the induced subgraph $\mathcal{G}_{\iota,b}^{K}$ serialized in DOT format that encodes static attributes$\rightarrow$behavior and behavior$\rightarrow$function relations with their corresponding evidence weights (\texttt{\{dot\_text\}}), iii) the descriptions (\texttt{\{movement\_text\}}) of the candidate set $\mathcal{C}_b^{K}$, and iv) contextual vessel attributes $\mathcal{V}_s(\iota)$ inferred from neighboring segments (\texttt{\{context\_vessels\}}).

These inputs jointly provide both structural and contextual evidence for inferring the most plausible behavior pattern. The prompt then compels the model to analyze $\mathcal{G}_{\iota,b}^{K}$ and the boundary context $(v_b^{-}, v_b^{+})$, select a single final movement ID from $\mathcal{C}_b^{K}$, and output a concise two-part rationale in a fixed, machine-readable schema. The first part, \emph{Graph Support}, cites key static attributes$\rightarrow$behavior edges and their weights that justify the selection, while the second part, \emph{Contextual Justification}, explains how the selected movement aligns with the boundary patterns and vessel context. The result is formatted as a triple-quoted block containing the selected ID, graph-based evidence, and contextual reasoning, which can be parsed into the rationale set $\mathcal{J}_{\iota}^{k,b}$ for downstream integration and interpretability.

%A.2.2
\begin{figure*}[t]
\centering
\begin{tcolorbox}[
  colback=gray!8, colframe=black!80,
  boxrule=0.5pt, arc=3mm,
  left=8pt, right=8pt, top=6pt, bottom=6pt,   % 给行号留出左边距
  title=Method Selector and Rationale,
  colbacktitle=gray!40, coltitle=black, fonttitle=\bfseries
]
\small
[\textbf{TASK}] You are an \textbf{expert in maritime spatio-temporal modeling and trajectory reconstruction}, responsible for evaluating and selecting the most suitable spatial function for accurate AIS trajectory imputation. \textbf{Please select the most suitable spatial function} for imputing missing latitude and longitude. \\
You need to adhere following requirements: 
\begin{itemize}[leftmargin=12pt,itemsep=1pt,topsep=1pt,parsep=0pt]
\item \textbf{Direction: }Which function has proven most reliable for similar kinematic patterns?
\item \textbf{Direction: }Does the function's underlying model (e.g., linear, curved) logically match the identified movement pattern (e.g., curved, straight)? 
\item \textbf{Direction: }Which function works best across different but related movement patterns?
\item \textbf{Direction: }How well does each function handle the specific speed/course/heading characteristics?
\item \textbf{Important: }When providing Statistical Support, ONLY discuss statistical evidence --- DO NOT mention any edge-weights and graph but you could turn it to statistical describe.
\item \textbf{Important: }Based on the calculation of weight proportions, all probabilities must be supported by evidence and cannot be arbitrarily fabricated. Each probability should be followed by its calculation process.
\item \textbf{Important: }Avoid repeating, movement pattern analysis --- focus on function execution quality.
\end{itemize}

[\textbf{INPUT}] You are given by: 
\begin{itemize}[leftmargin=12pt,itemsep=1pt,topsep=1pt,parsep=0pt]
\item \textbf{Induced subgraph in DOT (Interpret weights as association frequencies)}:
{\footnotesize\ttfamily \{dot\_text\}}
\item \textbf{Functions with detailed information}: 
{\footnotesize\ttfamily \{functions\_text\}}
\item \textbf{behavior pattern that may correspond to missing parts}: 
{\footnotesize\ttfamily \{movement\_text\}}
\item \textbf{AIS data of the neighboring segments}: 
{\footnotesize\ttfamily \{rows\_text\}}
\end{itemize}

[\textbf{OUTPUT}]
% \noindent \textbf{Return exactly in this format (no extra text):}
Please strictly follow the following format:\\
{\small\ttfamily
\verb|'''|\\
Selected Function ID: <ID> \\
Statistical Support: <Don't describe the edge weight, Don't describe the graph support but you could turn it to statistical describe.1. Introduce the probability that this function can solve the missing value problem corresponding to the behavior pattern(Based on the calculation of weight proportions, all probabilities must be supported by evidence and cannot be arbitrarily fabricated. Each probability should be followed by its calculation process.).  
2. Introduce the characteristics of this function and its degree of matching with the current context.> \\
Reasoning: <why this function technically fits the kinematic requirements> \\
\verb|'''|
}

\end{tcolorbox}
\vspace{-2mm}
\caption{Prompt of method selector and rationale.}
\label{fig:method-selector-prompt-exec}
\vspace{-2mm}
\end{figure*}

\begin{figure*}[t]
\centering
\begin{tcolorbox}[
  colback=gray!8, colframe=black!80,
  boxrule=0.5pt, arc=3mm,
  left=8pt, right=8pt, top=6pt, bottom=6pt,   % 给行号留出左边距
  title=Explanation Composer,
  colbacktitle=gray!40, coltitle=black, fonttitle=\bfseries
]
\small
[\textbf{TASK}]
You are an \textbf{expert in maritime behavior interpretation and regulatory reasoning}, specialized in translating computational decisions into human-understandable explanations for vessel trajectory analysis. Produce a \textbf{human-friendly explanation} for the chosen behavior and method. You need to adhere following requirements:
\begin{itemize}[leftmargin=12pt,itemsep=1pt,topsep=1pt,parsep=0pt]
  \item Do \textbf{NOT} mention any node IDs or labels such as ``Movement\_Pattern\_*'' or ``vessel\_*''.
  \item Refer \textbf{ONLY} to the concrete attributes and descriptions provided below.
\end{itemize}

[\textbf{INPUT}]
You are given by:
\begin{itemize}[leftmargin=12pt,itemsep=1pt,topsep=1pt,parsep=0pt]
  \item Induced subgraph in DOT:
  {\footnotesize\ttfamily \{dot\_text\}}
  \item Selected movement (expanded):
  {\footnotesize\ttfamily \{movement\_desc\}}
  \item Selected imputation method (expanded):
  {\footnotesize\ttfamily \{function\_desc\}}
  \item Contextual vessel attributes (expanded list):
  {\footnotesize\ttfamily \{vessels\_desc\_block\}}
  \item Contextual behavior pattern:
  {\footnotesize\ttfamily \{vessels\_behavior\_pattern\}}
\end{itemize}

% \medskip
% \noindent \textbf{Output exactly in the block below (no extra text):}
[\textbf{OUTPUT}]
Please strictly follow the following format:\\
{\small\ttfamily
\verb|'''|\\
Regulatory Rule Cue: <rule label + applicability + spatial anchor; leave Undetermined if insufficient> \\
Operational Protocol Rationale: <why this behavior is typical here; align with the spatial context; rule out key alternatives; do not mention IDs> \\
\verb|'''|
}

\end{tcolorbox}
% \vspace{-3mm}
\caption{Prompt of explanation composer.}
\label{fig:explanation-composer-prompt}
% \vspace{-3mm}
\end{figure*}

\subsubsection{Imputation Method Selection and Rationale} \label{prompt-MethodSelection}
Figure~\ref{fig:method-selector-prompt-exec} presents the prompt used by the Method Selector in Section~\ref{sec:MethodSelector}. 
Conditioned on the selected behavior $v_b^{*}$, the prompt supplies three structured inputs to the model: 
i) the induced subgraph $\mathcal{G}_{\iota,f}^{K}$ serialized in DOT, encoding behavior$\rightarrow$function relations and their association frequencies (\texttt{\{dot\_text\}}), 
ii) the candidate function set $\mathcal{C}_f^{K}$ with detailed metadata (\texttt{\{functions\_text\}}), 
and iii) the expanded description of the (already chosen) behavior $v_b^{*}$ to which the method must align (\texttt{\{movement\_text\}}); 
additionally, neighboring-segment AIS records are provided for execution-oriented context (\texttt{\{rows\_text\}}), while the selection itself emphasizes method quality over re-analysis of movement patterns. 
The prompt is instructed to evaluate $\mathcal{C}_f^{K}$ by jointly considering reliability on similar kinematic regimes, compatibility between the method’s underlying motion model (e.g., linear vs.\ curved) and $v_b^{*}$, cross-pattern robustness, and adaptability to observed speed/course/heading characteristics.

A strict output schema enforces machine-readability and interpretability: the model must return a triple-quoted block containing 
1) \texttt{Selected Function ID}, 
2) \texttt{Statistical Support}, which reports only statistical evidence (translated from association frequencies into probabilities) without mentioning graph structures or edge weights explicitly, each probability is accompanied by its calculation from weight proportions to prevent unverifiable claims—and
3) \texttt{Reasoning}, a technical justification of why the method matches the kinematic requirements of $v_b^{*}$. 
This schema ensures the selected method $v_f^{*}\in\mathcal{C}_f^{K}$ is both verifiable (via explicit probability construction from $\mathcal{G}_{\iota,f}^{K}$) and actionable (via the prescribed imputation procedure), while the resulting rationale is stored in $\mathcal{J}_{\iota}^{k,f}$ for downstream auditing and integration.

\begin{figure*}[t]
\centering
\begin{tcolorbox}[
  colback=gray!8, colframe=black!80,
  boxrule=0.5pt, arc=3mm,
  left=8pt, right=8pt, top=6pt, bottom=6pt,   % 给行号留出左边距
  title=De-redundancy Processor,
  colbacktitle=gray!40, coltitle=black, fonttitle=\bfseries
]
\small
[\textbf{TASK}]\\
You are an \textbf{expert in maritime knowledge consolidation and redundancy analysis}, specializing in detecting and merging semantically equivalent vessel behavior patterns and imputation functions. 
% Analyze maritime behavior patterns and spatial functions for redundancy and semantic similarity in one pass.
% \medskip
% \noindent \textbf{Context:}
% You are analyzing vessel behavior patterns and spatial imputation functions from AIS data.
You need to  adhere following requirements:\\
% \noindent \textbf{Analysis Criteria:}
For \textbf{Behavior Pattern}:
\begin{itemize}[leftmargin=14pt,itemsep=1pt,topsep=1pt,parsep=0pt]
  \item Exact duplicates: identical text
  \item Semantic equivalents: same meaning, different wording  
  \item Minor variations: slight wording differences
  \item Overly specific terms: very detailed descriptions
  \item Contextual synonyms: same meaning in maritime context
\end{itemize}

For \textbf{Imputation Function:}
\begin{itemize}[leftmargin=14pt,itemsep=1pt,topsep=1pt,parsep=0pt]
  \item Functional equivalence: different implementations but same mathematical function
  \item Algorithmic similarity: same core algorithm with minor variations
  \item Parameter differences: same logic with different parameter values
  \item Code restructuring: same functionality with different code structure
\end{itemize}

[\textbf{INPUT}]\\
% \medskip
% \noindent \textbf{Inputs:}
You are given by:
\begin{itemize}[leftmargin=12pt,itemsep=1pt,topsep=1pt,parsep=0pt]
  \item Behavior patterns to analyze:\\
  {\footnotesize\ttfamily \{vb\_data\_text\}}
  \item Spatial functions to analyze:\\
  {\footnotesize\ttfamily \{vf\_data\_text\}}
\end{itemize}

[\textbf{OUTPUT}]\\
Please strictly follow the following format: \\
For behavior patterns:
\begin{verbatim}
BEHAVIOR_REDUNDANCY:
[attribute_name]:
- <primary_term1> | [<redundant_term1>, <redundant_term2>]
- <primary_term2> | [<redundant_term3>, <redundant_term4>]
KEEP_UNIQUE: [<term1>, <term2>, <term3>]
\end{verbatim}

For spatial functions:

\begin{verbatim}
FUNCTION_REDUNDANCY:
- <primary_function_id_or_code> | [<redundant_function_id_or_code1>, <redundant_function_id_or_code2>]
- <primary_function_id_or_code> | [<redundant_function_id_or_code3>]
KEEP_UNIQUE: [<function_id_or_code1>, <function_id_or_code2>]
\end{verbatim}

\medskip
\noindent Focus on maritime behavior semantics and functional equivalence. Preserve meaningful distinctions.

\end{tcolorbox}
% \vspace{-2mm}
\caption{The prompt of de-redundancy processor.}
\label{fig:de-redundancy-processor-prompt}
% \vspace{-2mm}
\end{figure*}

%A.2.3

\subsubsection{Explanation Composer for Human-Friendly Explanation} \label{prompt-Explain}
Figure~\ref{fig:explanation-composer-prompt} presents the prompt used by the Explanation Composer in Section~\ref{sec:underlying-cause}. 
The prompt supplies a compact evidence view consisting of: 
i) the induced subgraph that binds the selected behavior and method with contextual entities, $\mathcal{G}_{d}\!\left[\mathcal{V}_s(\iota)\cup\{v_b^{*}\}\cup\{v_f^{*}\}\right]$ (serialized as DOT, \texttt{\{dot\_text\}}), 
ii) the expanded description of the selected behavior $v_b^{*}$ (\texttt{\{movement\_desc\}}), 
iii) the expanded description of the selected imputation method $v_f^{*}$ (\texttt{\{function\_desc\}}), 
iv) the static vessel attributes $\mathcal{V}_s(\iota)$ (\texttt{\{vessels\_desc\_block\}}), 
and v) the boundary behavior context $(v_b^{-},\,v_b^{+})$ that anchors spatial and operational intent (\texttt{\{vessels\_behavior\_pattern\}}). 
These inputs jointly instantiate the spatial context and the vessel profile needed for explanation.

The prompt is constrained not to reference internal node identifiers and to ground its explanation only in the concrete attributes provided by the inputs. It must produce a fixed, machine-readable output with two fields: 
1) \texttt{Regulatory Rule Cue}: a minimal yet sufficient statement of the governing rule in context (``rule label + applicability conditions + spatial anchor''), activated by $\sigma$ and $\mathcal{V}_s(\iota)$ and consistent with $v_b^{*}$; and 
2) \texttt{Operational Protocol Rationale}: a procedural account of why the observed behavior typically occurs here, aligned with the rule cue and $\sigma$, and briefly ruling out salient alternatives. 
The returned triple-quoted block (see Figure~\ref{fig:explanation-composer-prompt}) is recorded as $\mathcal{J}_{\iota}^{k,h}$ for downstream auditing and integration, ensuring that symbolic cues (rules and spatial anchors) and operational logic are preserved in a human-friendly, verifiable form without exposing graph-internal identifiers.

\begin{algorithm}[!t]
\caption{Knowledge-Driven Trajectory Imputation}
\small
\label{alg:kdt-impute}
\KwInput{AIS dataset $\mathcal{X}$; SD-KG $\mathcal{G}_d$; knowledge units of complete AIS data $\mathcal{U}_{d}$; thresholds $m$ and $K$.}
\KwOutput{Set of knowledge-supported imputation outcomes $\mathcal{R}$.}
$\mathcal{R}\gets \varnothing$ \\
\ForEach{vessel $\iota$ in $\mathcal{X}$}{
  $\langle \mathbf{S}_\iota^1,\dots,\mathbf{S}_\iota^K\rangle \gets \textsc{Partition}(\mathbf{X}_\iota; m)$\\
  $\mathbf{M}_\iota \gets \textsc{GetSegmentMask}(\langle \mathbf{S}_\iota^1,\dots,\mathbf{S}_\iota^K\rangle)$ \\
   $R_{\iota} \gets \langle \rangle$\\
  \ForEach{$k=1$ \KwTo $K$}{
    \uIf{$M_\iota^k=1$}{
       $R_{\iota} \gets R_{\iota} \| \varnothing$\\
        \textbf{continue}
    } 
    \bluecomment{Context Extraction}\\
    Obtain contextual knowledge units $u_{\iota}^{-},u_{\iota}^{+}$ from $\mathcal{U}_d$\\
    Infer $\mathcal{V}_s^{k}(\iota)$ from neighbors $(u_{\iota}^{-},u_{\iota}^{+})$\\
    Extract $(v_b^{-},v_b^{+})$ from $u_{\iota}^{-},u_{\iota}^{+}$\\
    \bluecomment{Behavior Estimator}\\
    $\mathcal{C}_b \gets \{\,v_b\in\mathcal{V}_b\mid \sum_{v\in \mathcal{V}_s^{k}(\iota)} w_{sb}(v,v_b)>0\,\}$\\
    Compute $\pi(v_b)$ by Eq.~\ref{eq:nodePrior};\quad $\mathcal{C}_b^{K}\gets \textsc{TopK}(\mathcal{C}_b,\pi,K_b)$\\
    $\mathcal{G}_{\iota,b}^{K}\gets \mathcal{G}_d[\mathcal{V}_s^{k}(\iota)\cup \mathcal{C}_b^{K}]$;\quad $\texttt{dot}_b\gets \textsc{SeriDOT}(\mathcal{G}_{\iota,b}^{K})$\\
    $\langle v_b^{*},\mathcal{J}_{\iota}^{k,b}\rangle \gets \textsc{Behavior\_Select\_Ration}(\texttt{dot}_b,\ v_b^{-},v_b^{+})$
    
    \bluecomment{Method Selector}\\
    $\mathcal{C}_f \gets \{\,v_f\in\mathcal{V}_f\mid w_{bf}(v_b^{*},v_f)>0\,\}$\\
    Compute $\pi(v_f)$;\quad $\mathcal{C}_f^{K}\gets \textsc{TopK}(\mathcal{C}_f,\pi,K_f)$\\
    $\mathcal{G}_{\iota,f}^{K}\gets \mathcal{G}_d[\{v_b^{*}\}\cup \mathcal{C}_f^{K}]$;\quad $\texttt{dot}_f\gets \textsc{SeriDOT}(\mathcal{G}_{\iota,f}^{K})$\\
    $\langle v_f^{*},\mathcal{J}_{\iota}^{k,f}\rangle \gets \textsc{Method\_Select\_Ration}(\texttt{dot}_f,\ v_b^{*})$
    
    \bluecomment{Execution}\\
    $\widehat{\mathbf{S}}_{\iota}^k \gets \textsc{Execute}(v_f^{*}.f,\ \mathbf{S}_{\iota}^k)$ 
    
    \bluecomment{Explanation Composer}\\
    $\mathcal{G}_{\iota}^{h} \gets \mathcal{G}_d[\mathcal{V}_s(\iota)\cup\{v_b^{*}\}\cup\{v_f^{*}\}]$\\
    $\texttt{dot}^{h}\gets \textsc{SeriDOT}(\mathcal{G}_{\iota}^{h})$\\
    $\mathcal{J}_{\iota}^{k,h} \gets \textsc{Exp\_Composer}(\texttt{dot}^{h},\ v_b^{*},\ v_f^{*},\ \mathcal{V}_s(\iota),\ v_b^{-},v_b^{+})$
    
    \bluecomment{Assemble Result}\\
    $R_{\iota} \gets R_{\iota} \| \big(\widehat{\mathbf{S}}_{\iota}^k,\; ((v_b^{*},\mathcal{J}_{\iota}^{k,b}),\ (v_f^{*},\mathcal{J}_{\iota}^{k,f}),\ \mathcal{J}_{\iota}^{k,h})\big)$\\
  }
  $\mathcal{R}\gets \mathcal{R}\cup \{R_{\iota}\}$
}
\KwRet $\mathcal{R}$
\end{algorithm}

\subsubsection{Overall Procedure of Knowledge-Driven Trajectory Imputation}
Algorithm~\ref{alg:kdt-impute} summarizes the dataset-level inference workflow that imputes all gaps using the SD-KG as a knowledge substrate. The procedure iterates over vessels and their segmented streams, retrieves boundary knowledge for each gap, selects a behavior and a method with graph-supported rationales, executes the method to reconstruct the missing segment, and composes a human-friendly explanation. 

Line~1 initializes the result accumulator $\mathcal{R}$. 
Lines~2--5 iterate over each vessel $\iota$, segment the AIS stream via \textsc{Partition} and build the completeness mask via \textsc{GetSegmentMask}; a per-vessel result sequence $R_\iota$ is created to preserve segment order.
Lines~6--9 traverse segments $k{=}1{:}K$. 
If $M_\iota^k{=}1$ (complete), append $\varnothing$ to $R_\iota$ and continue, preserving the index alignment between the original segmentation and the imputation outcomes.
Lines~10--13 fetch the left/right contextual knowledge units $(u_{\iota}^{-},u_{\iota}^{+})$ from the prebuilt collection $\mathcal{U}_d$ and infer the slowly varying static set $\mathcal{V}_s^{k}(\iota)$. 
Boundary behavior patterns $(v_b^{-},v_b^{+})$ are extracted from the neighboring units to anchor kinematic intent at the gap edges.
Lines~14--18 form the candidate set $\mathcal{C}_b$ through static$\rightarrow$behavior edges, compute the normalized log-additive prior $\pi(v_b)$ by Equation~\ref{eq:nodePrior}, and take the top-$K_b$ to obtain $\mathcal{C}_b^{K}$. 
An induced subgraph $\mathcal{G}_{\iota,b}^{K}$ over $\mathcal{V}_s^{k}(\iota)\cup \mathcal{C}_b^{K}$ is serialized to DOT (\textsc{SeriDOT}). 
\textsc{Behavior\_Select\_Ration} consumes this DOT together with $(v_b^{-},v_b^{+})$ and returns the selected behavior $v_b^{*}$ and its rationale $\mathcal{J}_{\iota}^{k,b}$ (graph support + contextual justification).
Lines~19--23 construct the candidate method set $\mathcal{C}_f$ via behavior$\rightarrow$function edges from $v_b^{*}$, score each candidate with $\pi(v_f)$ (same prior construction), and select the top-$K_f$ as $\mathcal{C}_f^{K}$. 
The induced subgraph $\mathcal{G}_{\iota,f}^{K}$ over $\{v_b^{*}\}\cup \mathcal{C}_f^{K}$ is serialized (DOT) and fed into \textsc{Method\_Select\_Ration}, which outputs the chosen method $v_f^{*}$ and its rationale $\mathcal{J}_{\iota}^{k,f}$.
Lines~24--25 run the executable function $v_f^{*}.f$ on the incomplete segment $\mathbf{S}_{\iota}^k$ to produce the reconstruction $\widehat{\mathbf{S}}_{\iota}^k$.
Lines~26--29 build a compact evidence view $\mathcal{G}_{\iota}^{h}$ over $\mathcal{V}_s(\iota)\cup\{v_b^{*}\}\cup\{v_f^{*}\}$, serialize it to DOT, and call \textsc{Exp\_Composer} to generate a human-friendly explanation $\mathcal{J}_{\iota}^{k,h}$ comprising the \emph{Regulatory Rule Cue} and the \emph{Operational Protocol Rationale}. 
(For consistency with other sections, this explanation is denoted $\mathcal{J}_{\iota}^{k,h}$ in the assembled tuple below.)
Lines~30--31 append to $R_\iota$ the triplet consisting of the reconstruction and the three rationales $\big(\widehat{\mathbf{S}}_{\iota}^k,\ ((v_b^{*},\mathcal{J}_{\iota}^{k,b}),\ (v_f^{*},\mathcal{J}_{\iota}^{k,f}),\ \mathcal{J}_{\iota}^{k,h})\big)$. 
Line~32 adds the per-vessel sequence $R_\iota$ to the global set $\mathcal{R}$. 
Line~33 returns $\mathcal{R}$, a vessel-indexed collection whose entries preserve segment order (with $\varnothing$ at complete positions) and, for every gap, provide (i) the reconstructed trajectory, (ii) graph-grounded rationales for behavior and method selection, and (iii) a human-readable explanation grounded in rules and operations.

\subsection{Workflow Management Layer}
\label{app:workflowmanagementlayer}

\subsubsection{De-redundancy Strategy}
\label{appendix:prompt-dedup}
Figure~\ref{fig:de-redundancy-processor-prompt} presents the prompt used by the De-redundancy Processor of SD-KG Construction Workflow Manager in Section~\ref{sec:build-manager}.
The prompt operates on two inputs: i) the raw set of behavior patterns $\mathcal{B}_{b}$ (\texttt{\{vb\_data\_text\}}) and ii) the raw set of imputation functions $\mathcal{B}_{f}$ (\texttt{\{vf\_data\_text\}}). 
For behaviors, the processor induces an equivalence/subsumption relation $\equiv_b$ over $\mathcal{B}_{b}$ based on the criteria enumerated in the prompt (exact duplicates, semantic equivalents, minor variations, overly specific terms, contextual synonyms). 
It then computes a canonicalization map $\pi_b:\mathcal{B}_{b}\!\to\!\widehat{\mathcal{B}}_b$ and corresponding clusters $C_b(\hat{b})=\{\,b\in\mathcal{B}_{b}:\pi_b(b)=\hat{b}\,\}$, where $\widehat{\mathcal{B}}_b$ denotes canonical representatives. 
Behavior terms that form singleton classes yet are not semantically collapsible are listed under \texttt{KEEP\_UNIQUE}. 
For functions, the processor forms an equivalence relation $\sim_f$ over $\mathcal{B}_{f}$ driven by functional equivalence, algorithmic similarity with minor variations, parameter-only differences, and code restructuring with identical semantics. 
It analogously derives a canonicalization map $\pi_f:\mathcal{B}_{f}\!\to\!\widehat{\mathcal{B}}_f$ and clusters $C_f(\hat{f})$, where $\widehat{\mathcal{B}}_f$ are canonical implementations retained for SD-KG insertion.

The prompt enforces a machine-readable output that mirrors these partitions. 
For behaviors, the block \texttt{BEHAVIOR\_REDUNDANCY} lists pairs ``\verb!primary term $|$ [redundant terms]!'' grouped by attribute (e.g., maneuver, route, intent), with residuals reported in \texttt{KEEP\_UNIQUE}. 
For functions, \texttt{FUNCTION\_REDUNDANCY} lists ``\verb!primary! \verb!function id/code $|$ [redundant ids/codes]!'', with non-mergeable items collected under \texttt{KEEP\_UNIQUE}.
This schema yields two canonical sets $(\widehat{\mathcal{B}}_b,\widehat{\mathcal{B}}_f)$ and explicit many-to-one provenance mappings $(\pi_b,\pi_f)$, ensuring that SD-KG construction remains compact while preserving meaningful distinctions required for downstream reasoning and execution.

\begin{algorithm}[!t]
\caption{SD-KG Construction Workflow Manager}
\small
\label{alg:sdk-workflow}
\KwInput{AIS dataset $\chi$; batch size $b$; thresholds $\varrho_c$, $\varrho_r$, $\varrho_f$, $m$.}
\KwOutput{SD-KG $\mathcal{G}_d$; knowledge units $\mathcal{U}_{d}$.}

\bluecomment{Build Job Stack (Stack-Based Scheduler)}\\
$\mathcal{S}_c, \mathcal{S}_d \gets \textsc{EmptyStack}(), \textsc{EmptyStack}()$ \\
\ForEach{vessel $\iota$}{
  $\langle \mathbf{S}_\iota^1,\dots,\mathbf{S}_\iota^K\rangle \gets \textsc{Partition}(\mathbf{X}_\iota; m)$\\
  \ForEach{segment $\mathbf{S}_\iota^k$ sorted by timestamp}{
    \textsc{Push}$(\mathcal{S}_c,\ ( \iota,k,\mathbf{S}_\iota^k, 0))$
  }
}
\bluecomment{Parallel Knowledge Unit Extraction}\\
\While{$\mathcal{S}_c \neq \varnothing$}{
  $\mathcal{B} \gets \textsc{PopBatch}(\mathcal{S}_c,\ b)$ \\
  \bluecomment{Launch Parallel Extraction Jobs}\\
  \ParFor{$(\iota,k,\mathbf{S}_\iota^k,c)\in\mathcal{B}$}{
  $u_{\iota}^k \gets \textsc{ExtractKU}(\mathbf{S}_\iota^k,\varrho_r, \varrho_f)$
  }
  \bluecomment{Anomaly Guard}\\
  $\mathcal{B}_u =\varnothing$\\
  \ForEach{$u_{\iota}^k \in \{\,u_{\iota}^k \mid (\iota,k)\in\mathcal{B}\,\}$}{
    \uIf{$\textsc{Anomaly\_Detect}(u_{\iota}^k)$}{
      \uIf{$c < \varrho_c$}{
        \textsc{Push}$(\mathcal{S}_c,\ ( \iota,k,\mathbf{S}_\iota^k, c+1))$
      }
    }
    \Else{
      $\mathcal{B}_u = \mathcal{B}_u \cup \{u_{\iota}^k\}$
    }
  }
  \textsc{Push}$(\mathcal{S}_d,\ \mathcal{B}_u)$\\
}
\bluecomment{Parallel De-redundancy Processor}\\
$\mathcal{U}_d \gets \varnothing$\\
\While{$\mathcal{S}_d \neq \varnothing$}{
  $\mathcal{B} \gets \textsc{PopBatch}(\mathcal{S}_d,\ b)$ \\
  \bluecomment{Launch Parallel De-redundancy Jobs}\\
  \ParFor{$\mathcal{B}_u \in\mathcal{B}$}{
  $\mathcal{B}_{u}^d \gets \textsc{Deredund\_Processor}(\mathcal{B}_u)$
  }
  $\mathcal{U}_d \gets \mathcal{U}_d \cup \{\,\mathcal{B}^d_{u} \mid \mathcal{B}_{u} \in\mathcal{B}\,\}$\\
}
\bluecomment{Write to SD-KG}\\
\ForEach{$(v_s,v_b,v_f) \in \mathcal{U}_{d}$}{
    $\mathcal{V}_s, \mathcal{V}_b, \mathcal{V}_f \gets \mathcal{V}_s \cup v_s, \mathcal{V}_b \cup \{v_b\}, \mathcal{V}_f \cup \{v_f\}$\\
    $\mathcal{E}_{sb},\mathcal{E}_{bf} \gets \textsc{EdgeInc} ((v_s,v_b,v_f), \mathcal{E}_{sb},\mathcal{E}_{bf})$\\
}
\KwRet $\big(\mathcal{G}_d,\ \mathcal{U}_{d}\big)$
\end{algorithm}

\subsubsection{Overall Procedure of SD-KG Construction Workflow Manager}
Algorithm~\ref{alg:sdk-workflow} orchestrates knowledge extraction with a two-stage parallel workflow (extraction $\rightarrow$ de-redundancy) connected by stack-based scheduling and batch barriers.
Lines 1--6 initialize the workflow by constructing two stacks: the compute stack $\mathcal{S}_c$ and the de-duplication stack $\mathcal{S}_d$. All minimal segments $\mathbf{S}_\iota^k$ are pushed to $\mathcal{S}_c$ in timestamp order, each as a job tuple $(\iota,k,\mathbf{S}_\iota^k,c)$ with an initial retry counter $c{=}0$.
Lines 7--12 launch batched parallel extraction. While $\mathcal{S}_c$ is non-empty, a batch of at most $b$ jobs is popped and executed concurrently (\textsc{ParFor}), where each job calls \textsc{ExtractKU} to produce a knowledge unit $u_{\iota}^k$ for segment $\mathbf{S}_\iota^k$. A synchronization barrier follows each batch to ensure consistent collection of results.
Lines 13--21 perform anomaly detection and bounded retry. 
Each extracted unit is examined by \textsc{Anomaly\_Detect}.
If an anomaly is found and the retry counter satisfies $c<\varrho_c$, the job is re-queued to $\mathcal{S}_c$ with counter incremented to $c{+}1$.
Otherwise, the unit is admitted into the valid batch $\mathcal{B}_u$.
Jobs that exceed $\varrho_c$ retries are quarantined.
The validated batch $\mathcal{B}_u$ is then pushed onto $\mathcal{S}_d$ for de-redundancy.
Lines 22--29 execute the parallel de-redundancy stage.
Each batch popped from $\mathcal{S}_d$ is processed in parallel by \textsc{Deredund\_Processor}, which merges semantically equivalent behavior tokens and function implementations to produce de-duplicated results $\mathcal{B}_u^{d}$.
All cleaned units are accumulated into $\mathcal{U}_d$, enabling scalable consolidation independent of extraction.
Lines 30--33 commit the consolidated results to the SD-KG. 
Each $(v_s,v_b,v_f)\in\mathcal{U}_d$ is inserted or reused within $\mathcal{V}_s,\mathcal{V}_b,\mathcal{V}_f$, and the corresponding edges $(v_s,v_b)$ and $(v_b,v_f)$ are updated via \textsc{EdgeInc}. 
Each per-vessel sequence $\mathcal{U}_{d,\iota}$ is maintained in temporal order to align with segment index $k$.
Line 34 returns the updated SD-KG $\mathcal{G}_d$ and the complete set of de-duplicated knowledge units $\mathcal{U}_d$.

Batch size $b$ constrains concurrent jobs to balance throughput and stability; the dual-stack scheduling pattern decouples extraction and consolidation, forming a pipeline that overlaps compute with integration. The retry threshold $\varrho_c$ safeguards against unstable segments, while the refinement bound $\varrho_r$ inside \textsc{ExtractKU} ensures quality control of generated imputation methods.

\begin{algorithm}[!t]

\caption{Trajectory Imputation Workflow Manager}

\small

\label{alg:impute-workflow}

\KwInput{AIS dataset $\mathcal{X}$; SD-KG $\mathcal{G}_d$; knowledge units of complete AIS data $\mathcal{U}_{d}$; batch size $b$; retry limit $\varrho_i$;  thresholds $m$ and $K$.}
\KwOutput{Reconstructed trajectories with explanations $\mathcal{R}$.}

\bluecomment{Build Job Stack (Stack-Based Scheduler)}\\
$\mathcal{S}_i \gets \textsc{EmptyStack}()$\\
\ForEach{vessel $\iota$}{
  $\langle \mathbf{S}_\iota^1,\dots,\mathbf{S}_\iota^K\rangle \gets \textsc{Partition}(\mathbf{X}_\iota; m)$\\
  \ForEach{gap segment $\mathbf{S}_\iota^k$ sorted by timestamp}{
    \textsc{Push}$(\mathcal{S}_i, ( \iota,k,\mathbf{S}_\iota^k,0))$
  }
}
$\mathcal{R}\gets\varnothing$\\
\While{$\mathcal{S}_i \neq \varnothing$}{
   $\mathcal{B} \gets \textsc{PopBatch}(\mathcal{S}_i,\ b)$\\ 
    \bluecomment{Launch Parallel Imputation Jobs}\\  \ParFor{$( \iota,k,\mathbf{S}_\iota^k,c)\in\mathcal{B}$}{
    $\mathcal{R}^k_{\iota} \gets \textsc{Traj\_Imputation}(\mathcal{G}_d, \mathcal{U}_d, ( \iota, k,\mathbf{S}_\iota^k,c),K)$\\
  }
  \bluecomment{Anomaly Guard}\\
  \ForEach{$\mathcal{R}^k_{\iota} \in \{\,\mathcal{R}^k_{\iota} \mid (\iota,k)\in\mathcal{B}\,\}$}{
    \uIf{$\textsc{Anomaly\_Detect}(\mathcal{R}^k_{\iota})$}{
      \uIf{$c < \varrho_i$}{
        \textsc{Push}$(\mathcal{S}_i,\ ( \iota,k,\mathbf{S}_\iota^k, c+1))$
      }
    }
    \Else{
      $\mathcal{R} = \mathcal{R} \cup \{\mathcal{R}^k_{\iota}\}$
    }
  }
}
\KwRet $\mathcal{R}$
\end{algorithm}

\subsubsection{Overall Procedure of Trajectory Imputation Workflow Manager}
\label{sec:alg-impute}
Algorithm~\ref{alg:impute-workflow} summarizes the dataset-level batch-parallel pipel\-ine that reconstructs all gaps using the SD-KG. 
Lines 1--6 build the imputation job stack $\mathcal{S}_i$: for each vessel, the stream is segmented (\textsc{Partition}), and every gap segment is timestamp-sorted and pushed as a tuple $(\iota,k,\mathbf{S}_\iota^k,0)$ with an initial retry counter. 
Line 7 initializes the global result set $\mathcal{R}$. 
Lines 8--12 enter the scheduling loop: each iteration pops up to $b$ jobs to form a micro-batch $\mathcal{B}$ and launches parallel workers, each invoking the one-call pipeline \textsc{Traj\_Imputation}$(\mathcal{G}_d,\mathcal{U}_d,(\iota,k,\mathbf{S}_\iota^k,c),K)$, which internally performs behavior selection, method selection, execution, and explanation (Section~\ref{sec:InterpretableImputation}). 
Lines 13--19 implement the \emph{Anomaly Guard}: for every per-segment result $\mathcal{R}^k_\iota$ in $\mathcal{B}$, \textsc{Anomaly\_Detect} filters invalid outcomes (e.g., empty response or non-executable selections). 
Failed jobs are re-queued with an incremented retry counter if $c<\varrho_i$; otherwise they are dropped from the current pass (cf. Section~\ref{sec:impute-manager} on logging). Valid results are aggregated into $\mathcal{R}$. 
Line 20 returns $\mathcal{R}$, which collects, for each imputed gap, the reconstructed trajectory together with graph-grounded rationales and a human-friendly explanation, preserving segment order at the vessel level.

% \section{Comprehensive Experimental Configuration and Supplementary Results}
\section{Experimental Configuration}
\label{app:Detailed Experimental Setting}
\subsection{Datasets}
We use two AIS datasets: AIS-DK from the Danish Maritime Authority and AIS-US from the National Oceanic and Atmospheric Administration. \textbf{AIS-DK} covers Danish waters in March 2024, including major routes in the Baltic and North Seas; it contains 10,000 vessel sequences and 2,000,000 records, with an average sequence duration of 0.5 hours across 348 vessels. \textbf{AIS-US} covers U.S. coastal waters in April 2024 with dense traffic near major ports; it contains 10,000 vessel sequences and 2,000,000 records, with an average sequence duration of 2.8 hours across 4,723 vessels. Both datasets include diverse vessel types (e.g., Cargo, Tanker, Passenger), providing representative coverage for evaluation. To construct these datasets, we first downloaded the raw AIS data from the official sources and performed data cleaning to remove abnormal or incomplete records. We then partitioned each vessel trajectory---March 2024 for AIS-DK and April 2024 for AIS-US---into fixed-length segments of 200 records, and uniformly sampled 10,000 segments to build the evaluation corpus.

\subsection{Hyperparameters and Implementations}
% Hyperparameters and implementations are standardized as follows. 
Lin-ITP, Akima Spline, and Kalman Filter are implemented in Python using pandas and pykalman. 
% Lin-ITP has no tunable hyperparameters; the Akima spline uses a smoothing factor constrained between 0.0 (minimum) and 1.0 (maximum), with all other settings left at their defaults; and the Kalman filter also uses default hyperparameters.
MH-GIN~\cite{liu2025mhgin}, Multi-task AIS~\cite{nguyen2018multi}, and KAMEL~\cite{musleh2023kamel} follow the authors' default configurations. 
GLM-4.5, GLM-4.5-air, Qwen-Plus (snapshot 0112), and Qwen-flash (snapshot 2025-07-28) are accessed through the Aliyun Bailian API~\cite{bailian2025} with thinking mode enabled.

% To ensure data quality and representativeness, we construct these datasets from raw AIS streams using a four stage processing pipeline. First, we remove records with any missing attributes. Second, we discard entries with invalid fields such as out of range longitude or latitude and illegal navigation status codes. Third, we enforce a minimum inter record interval to reduce bursty reporting while preserving continuity, set to 80 s for AIS-DK and 300 s for AIS-US. Fourth, we partition each vessel trajectory into fixed length segments of 200 records and uniformly sample 1{,}000 segments to form the evaluation corpus.
% To ensure data quality and representativeness, we construct these datasets from raw AIS streams using a four stage processing pipeline. First, we remove records with any missing attributes. Second, we discard entries with invalid fields such as out of range longitude or latitude and illegal navigation status codes. Third, we enforce a minimum inter record interval to reduce bursty reporting while preserving continuity, set to 80 s for AIS-DK and 300 s for AIS-US. Fourth, we partition each vessel trajectory into fixed length segments of 200 records and uniformly sample 1{,}000 segments to form the evaluation corpus.

\begin{figure*}[t]
\centering
\begin{tcolorbox}[
  colback=gray!8, colframe=black!80,
  boxrule=0.5pt, arc=3mm,
  left=8pt, right=8pt, top=6pt, bottom=6pt,   % 给行号留出左边距
  title=Vessel Trajectory Imputation,
  colbacktitle=gray!40, coltitle=black, fonttitle=\bfseries
]
\small
[\textbf{TASK}]\\
You are a professional maritime data analyst. Please predict the missing trajectory segment based on the known vessel trajectory points.\\
You must strictly adhere to the following requirements:
\begin{itemize}[leftmargin=14pt,itemsep=1pt,topsep=1pt,parsep=0pt]
  \item Output strict JSON array format containing \textbf{\{missing\_length\}} coordinate points
  \item Each coordinate point in [longitude, latitude] format, keep 6 decimal places
  \item Ensure smooth and continuous trajectory that follows vessel movement patterns
  \item Consider temporal continuity and generate reasonable intermediate trajectory
  \item Do not add any explanatory text, only output JSON array
\end{itemize}

\medskip
[\textbf{INPUT}]
\begin{itemize}[leftmargin=12pt,itemsep=1pt,topsep=1pt,parsep=0pt]
  \item Previous trajectory point (last known point):
  {\footnotesize\ttfamily \{prev\_str\}}
  \item Next trajectory point (first known point after gap):
  {\footnotesize\ttfamily \{next\_str\}}
\end{itemize}

\medskip
[\textbf{OUTPUT}]\\
Predict the intermediate missing \textbf{\{missing\_length\}} trajectory points between these two points.

\medskip
[\textbf{EXAMPLE}]\\
\verb|'''|\\
{\small\ttfamily
[[121.123456, 31.234567], [121.124567, 31.235678], ...]
}\\
\verb|'''|

\end{tcolorbox}
\caption{Prompt of general large language model for vessel trajectory imputation.}
\label{fig:general-LLM-prompt}
\end{figure*}
\subsection{Evaluation Metrics}
\label{sec:appendix_evaluation_metrics}
We evaluate only timestamps inside minimal segments $\mathbf{S}_{\iota}^k$ that are imputed. A segment-level mask $M_{\iota}^k\in\{0,1\}$ identifies real gaps ($M_{\iota}^k=0$) or fully observed segments ($M_{\iota}^k=1$) from which we create synthetic gaps (see Definition~\ref{def-segmentmask}). Inside each selected segment, an internal mask $\widetilde{\mathbf m}_{\iota}^k\in\{0,1\}^m$ flags the timestamps that require evaluation.
Let the global index set be
\begin{equation}
\mathcal{I}=\big\{(\iota,k,j)\,:\,\widetilde m_{\iota,j}^k=1\big\},\quad |\mathcal{I}|~\text{its size},
\end{equation}
and denote ground-truth and imputed coordinates at index $(\iota,k,j)$ by
$(\lambda_{\iota,k,j},\phi_{\iota,k,j})$ and $(\hat\lambda_{\iota,k,j},\hat\phi_{\iota,k,j})$, respectively.

We adopt axis-wise Mean Absolute Error (MAE) to capture the average magnitude of errors robustly and Root Mean Squared Error (RMSE) to emphasize large deviations; both are reported in degrees. The formulas are:
\begin{align}
\mathrm{MAE}_\lambda &= \frac{1}{|\mathcal{I}|}\sum_{(\iota,k,j)\in\mathcal{I}}
\left|\hat \lambda_{\iota,k,j}-\lambda_{\iota,k,j}\right|,\\
\mathrm{MAE}_\phi &= \frac{1}{|\mathcal{I}|}\sum_{(\iota,k,j)\in\mathcal{I}}
\left|\hat \phi_{\iota,k,j}-\phi_{\iota,k,j}\right|,\\
\mathrm{RMSE}_\lambda &= \left(\frac{1}{|\mathcal{I}|} 
\sum_{(\iota,k,j)\in\mathcal{I}}
\left(\hat \lambda_{\iota,k,j} - \lambda_{\iota,k,j}\right)^2\right)^{1/2},\\
\mathrm{RMSE}_\phi &= \left(\frac{1}{|\mathcal{I}|}\sum_{(\iota,k,j)\in\mathcal{I}}
\left(\hat \phi_{\iota,k,j}-\phi_{\iota,k,j}\right)^2\right)^{1/2}
\end{align}

We use the Haversine geodesic distance as the joint spatial error and report it in kilometers. The formulas are:
\begin{align}
d_{\iota,k,j}  = 2R\,\arcsin\! & (\sqrt{\operatorname{hav}(\Delta\phi_{\iota,k,j}^r)
+\cos\phi^{r}_{\iota,k,j}\cos\hat\phi^{r}_{\iota,k,j}\,
\operatorname{hav}(\Delta\lambda_{\iota,k,j}^r)}),\\
\mathrm{MHD} & =\frac{1}{|\mathcal{I}|}\sum_{(\iota,k,j)\in\mathcal{I}} d_{\iota,k,j},
\end{align}
where $\operatorname{hav}(x)=\sin^2(x/2)$; $R$ is Earth’s mean radius (6371 km);
$\Delta\phi_{\iota,k,j}^r$ and $\Delta\lambda_{\iota,k,j}^r$ are the latitude and longitude differences
in radians; $\phi^{r}_{\iota,k,j}$ and $\hat\phi^{r}_{\iota,k,j}$ are the ground-truth and imputed latitudes.

\subsection{Baseline Methods}

We compare \textsf{VISTA} with three classes of baselines: rule-based trajectory imputation, deep learning-based trajectory imputation, and LLM-based trajectory imputation.

For rule-based trajectory imputation, we include \textbf{Lin-ITP}~\cite{widyantara_improvement_2023}, \textbf{Akima Spline}~\cite{zaman_interpolation-based_2023}, and \textbf{Kalman Filter}~\cite{wang_kinematic_2023}. 1) Lin-ITP linearly interpolates longitude and latitude between boundary observations, implicitly assuming constant velocity and heading, and is simple and fast. 2) Akima Spline constructs a piecewise cubic polynomial using local slopes estimated from neighboring points, avoiding oscillations and overshooting common in standard cubic splines while maintaining smooth curvature continuity. It performs better on irregular or nonuniform motion patterns. 3) Kalman Filter treats motion as a linear Gaussian state space with constant velocity dynamics; forward filtering with backward smoothing recovers latent states and positions and is most effective when kinematics are near-linear under moderate Gaussian noise.

For deep learning-based trajectory imputation, we include \textbf{MH-GIN}~\cite{liu2025mhgin} and \textbf{Multi-task AIS}~\cite{nguyen2018multi}. MH-GIN is a multi-scale heterogeneous graph imputation network for AIS streams: it extracts multi-scale temporal features per attribute while preserving heterogeneous update rates, then constructs a multi-scale heterogeneous graph to capture cross-attribute dependencies and propagates information to fill missing values. Multi-task AIS is a recurrent framework with latent variable modeling and an embedding of AIS messages designed for high-volume, noisy, and irregularly sampled data; it jointly supports trajectory reconstruction, anomaly detection, and vessel type identification.

\label{appendix:general-LLM-prompt}
For LLM-based trajectory imputation, we include \textbf{KAMEL}~\cite{musleh2023kamel}, \textbf{Qwen-plus-th}, \textbf{Qwen-flash-th}, \textbf{GLM-4.5-th}, and \textbf{GLM-4.5-air-th}. KAMEL casts trajectory completion as a missing-word problem with spatially aware tokenization and multi-point masked infilling. Beyond KAMEL, to probe the out-of-the-box capability of general LLMs on trajectory imputation without any task-specific design, we adopt two state-of-the-art families, Qwen~\cite{team_Qwen3_2025} and GLM 4.5~\cite{zeng2025glm}. For each family we include a lightweight and a full-capacity variant, and we enable thinking mode~\cite{Alibaba_Deep_2025} throughout (denoted by the ``-th'' suffix). This yields two paired comparisons under a unified setup: Qwen-plus-th vs Qwen-flash-th for capacity scaling, and GLM-4.5-air-th vs GLM-4.5-th for the trade-off between latency and accuracy.

\subsection{Prompt of General Large Language Model for Vessel Trajectory Imputation}
Figure~\ref{fig:general-LLM-prompt} presents the baseline prompt used to elicit intermediate points for a missing trajectory segment from a general-purpose LLM. The prompt conditions the model on two boundary observations, previous point ${p}^{-}=[\lambda^{-},\phi^{-}]$ (\texttt{\{prev\_str\}}) and next (first post-gap) point ${p}^{+}=[\lambda^{+},\phi^{+}]$ (\texttt{\{next\_str\}}), and a target count $m$ of missing samples (\texttt{\{missing\_length\}}). The model must output a strict JSON array of $m$ coordinates
$\big([\lambda_1,\phi_1],\ldots,[\lambda_m,\phi_m]\big)$,
each in \texttt{[longitude, latitude]} with six decimal places, without accompanying text. The constraints emphasize (i) smooth and continuous spatial progression consistent with vessel motion heuristics and (ii) temporal continuity for plausible interpolation between ${p}^{-}$ and ${p}^{+}$. The example block in Figure~\ref{fig:general-LLM-prompt} specifies the exact output schema to ensure deterministic parsing and fair, method-agnostic evaluation of baseline LLMs.

% \subsection{Prompt of LLM-based Baselines}

\section{Case Study}
\label{app:casestudy}

\begin{figure*}[!htbp]
\centering
\includegraphics[width=1\linewidth]{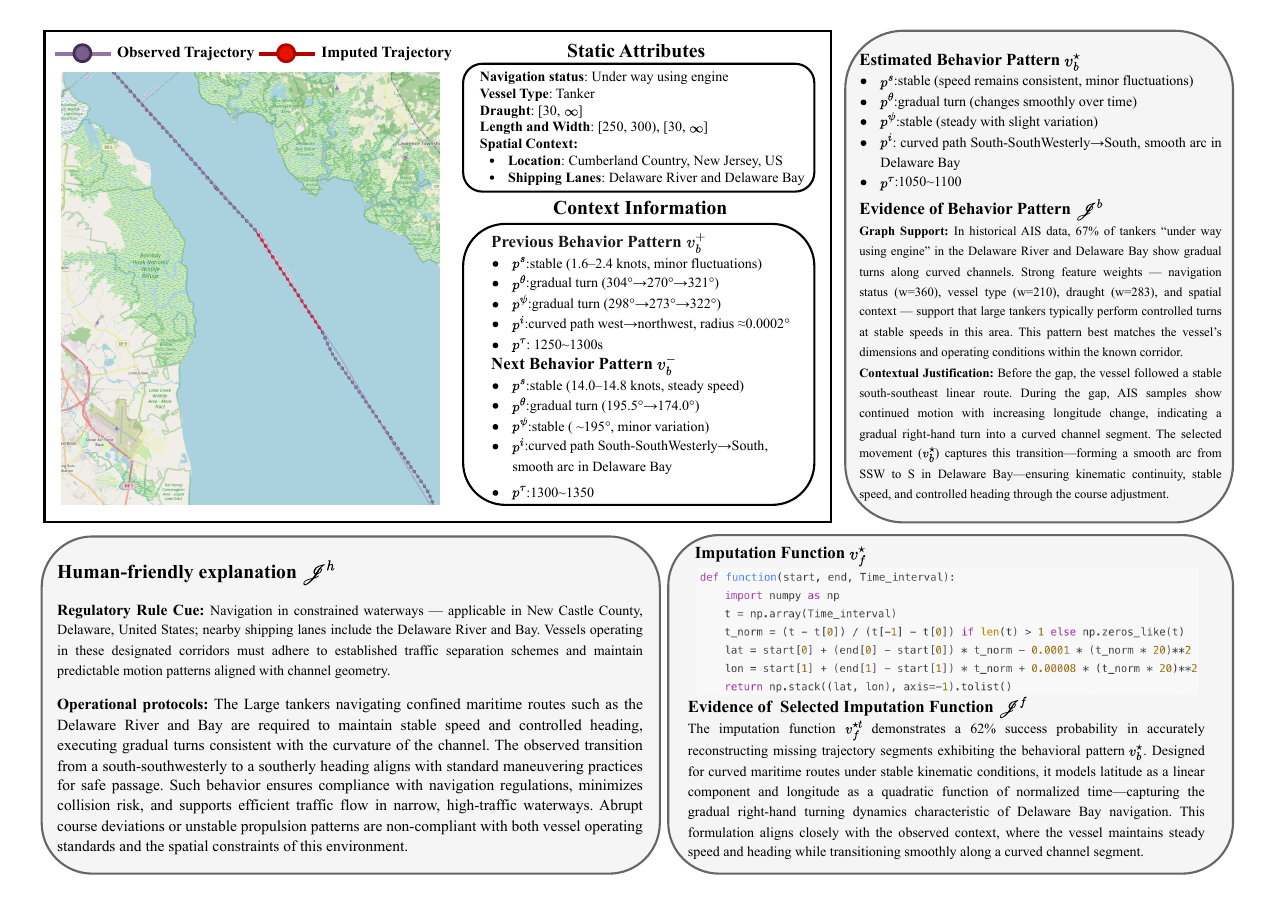}
\vspace{-6mm}
\caption{Case study of \textsf{VISTA}.}
\vspace{-3.5mm}
\label{fig:Case Study}
\end{figure*}

As shown in Figure~\ref{fig:Case Study}, this case study illustrates \textsf{VISTA}'s knowledge-driven imputation in constrained waterways near Delaware Bay. A tanker with missing AIS data (1050--1100 s) was analyzed using structured data-derived knowledge and LLM reasoning. Based on preceding and following motion cues, \textsf{VISTA} identified a stable-speed, gradual port-side turn pattern, consistent with 67\% of similar tankers ``under way using engine'' in this region. The selected spatial imputation function reconstructs the missing path as a smooth curved arc, preserving kinematic continuity.
The human-friendly explanation links the maneuver to navigation in constrained waterways in New Castle County, Delaware, within designated corridors (Delaware River and Bay) governed by traffic separation schemes that require predictable motion aligned with channel geometry, and to operational protocols for large tankers that maintain stable speed and controlled heading, executing gradual turns (from south-southwesterly to southerly) for safe passage.
The resulting trajectory aligns with maritime norms and environmental constraints, showing that the vessel's behavior is compliant rather than anomalous, demonstrating how \textsf{VISTA} integrates data-driven evidence and domain reasoning for interpretable trajectory reconstruction.

\end{document}